\documentclass[12pt,a4paper]{article}
\usepackage[format=hang]{caption}
\usepackage{epsfig,amssymb,amsmath,graphicx,subcaption,verbatim,hyperref}
\newcommand{\be}{\begin{equation}}
\newcommand{\ee}{\end{equation}}

\newcommand{\bpi}{\mbox{\boldmath $\pi$}}
\newcommand{\pauli}{\mbox{\boldmath $\tau$}}
\newcommand{\half}{1/2}

\newcommand{\de}{\mathrm{d}}

\font\mybb=msbm10 at 11pt
\def\bb#1{\hbox{\mybb#1}}

\def\bR {\bb{R}}

\newcommand{\news}{\setcounter{equation}{0}}
\def\bea{\begin{eqnarray}}
\def\eea{\end{eqnarray}}
\numberwithin{equation}{section}

\begin{document} 
\title{\vskip -70pt
\begin{flushright}
{\normalsize DAMTP-2012-62} \\
\end{flushright}
\vskip 60pt
{\bf {\LARGE Skyrmions up to Baryon Number 108}}\\[20pt]}
\author{\bf {\Large D.T.J. Feist\footnote{D.Feist@damtp.cam.ac.uk}, P.H.C. Lau\footnote{P.H.C.Lau@damtp.cam.ac.uk} and N.S. Manton\footnote{N.S.Manton@damtp.cam.ac.uk}} \\[25pt]
Department of Applied Mathematics and Theoretical Physics\\
University of Cambridge\\
Wilberforce Road, Cambridge CB3 0WA, England}

\date{October 2012}
\maketitle
\vskip 40pt
 
\begin{abstract}
The Skyrme crystal is built up of repeating units similar to the cubic
Skyrmion of baryon number 4. Using this as guide, we construct new Skyrmion
solutions in the massive pion case, with various baryon numbers up to 
108. Most of our solutions resemble chunks of the Skyrme crystal. They 
are constructed using a multi-layer version of the rational map ansatz
to create initial configurations, which are then relaxed numerically
to find the energy minima. The coefficients of the rational maps are 
found by a geometrical construction related to the Skyrme crystal 
structure. We find some further solutions by numerical relaxation 
of clusters composed of baryon number 4 Skyrmions.
\end{abstract}

\newpage



\section{Introduction}\news

The Skyrme model was proposed in 1961 by THR Skyrme as a model for nuclear physics \cite{Sk,book,BR}. It is 
a nonlinear field theory of pions admitting solitonic solutions 
called Skyrmions. Skyrmions model atomic nuclei, and have a conserved,
integer-valued topological charge $B$, which is interpreted as the 
baryon number, or mass number, of a nucleus.

The Skyrme field $U(x)$ is an $SU(2)$-valued scalar field. It can be written as
\be
U(x) = \sigma(x) \, {\boldsymbol I} + i \bpi(x) \cdot \pauli
\ee
with pion fields $\bpi = (\pi_1, \pi_2, \pi_3)$ and sigma field
$\sigma$. Here $x = (t,{\mathbf x}) = (x_0,x_1,x_2,x_3)$ and $\pauli$ are the 
Pauli matrices. $\sigma$ and $\bpi$ are not independent as
$UU^{\dagger} = (\sigma^2 + \bpi \cdot \bpi){\boldsymbol I} =
{\boldsymbol I}$.

The Lagrangian, in ``Skyrme units'', is 
\be
L=\frac{1}{12 \pi^2} \int \left\{-\frac{1}{2}\mbox{Tr}(R_\mu R^\mu)
+\frac{1}{16}\mbox{Tr}([R_\mu,R_\nu][R^\mu,R^\nu])
-m^2\mbox{Tr}({\boldsymbol I}-U)\right\} \, \de^3x \,,
\label{skylag}
\ee
where $R_{\mu}=(\partial_{\mu} U)U^\dagger$ is the right current.
The energy unit is roughly
$700\;\mathrm{MeV}$ and the length unit is roughly $1\;\mathrm{fm}$. The 
parameter $m$ is the pion mass in Skyrme units. Using the 
physical pion mass one finds $m \simeq 0.5$ \cite{AN}, but this is sensitive
to the length unit. It has been found that if one calibrates the model 
to real spinning nucleons, and some larger nuclei like 
Carbon-12 in its ground and excited states, a value of $m=1$ gives a better 
fit \cite{BS11,BKS,BMSW}. Therefore $m=1$ is used in this paper. 

For a static field $U({\mathbf x})$, the
energy depends only on $U$ and its spatial derivatives encoded in
the spatial current $R_i$, and is
\be
E= \frac{1}{12 \pi^2} \int \left\{-\frac{1}{2}\mbox{Tr}(R_iR_i)-\frac{1}{16}
\mbox{Tr}([R_i,R_j][R_i,R_j])+m^2\mbox{Tr}({\boldsymbol I}-U)\right\} 
\, \de^3x \,.
\label{skyenergy}
\ee
$U = {\boldsymbol I}$ is the vacuum of the theory. For a finite energy
field configuration, it is necessary that $U \rightarrow \boldsymbol
I$ as $|{\bf x}| \rightarrow \infty$, hence $\sigma \to 1$ and 
$\bpi \to {\bf 0}$. 

The baryon number $B$ is the topological degree of the
map $U: \bR ^3 \to SU(2)$, which is well-defined because
$U \to {\boldsymbol I}$ at spatial infinity. $B$ can be defined as the 
integral over $\bR ^3$ of the baryon density
\be
{\cal B} = -\frac{1}{24\pi^2}\epsilon_{ijk}{\rm Tr}(R_iR_jR_k) \,. 
\label{eq:bardens}
\ee
The minimum energy field configurations for each $B$ are the true
Skyrmions. For all but the smallest baryon numbers one finds further
configurations that are local minima of the energy, with very similar
energies, and these are also called Skyrmions. 

The Skyrme model with massless pions ($m=0$) has been studied intensively, and
Skyrmions with all baryon numbers up to $22$ have been found
\cite{BS3c}. Beyond baryon numbers $1$ and $2$, these Skyrmions have 
a polyhedral shell-like structure, surrounding a hollow region of 
small baryon density whose volume increases proportionally to the 
baryon number. This does not model real nuclei well. In recent 
years, the Skyrme model has been studied with a pion mass $m$ around 1, and 
Skyrmions with selected baryon numbers up to $B=32$ have been 
found \cite{BS11,BS10,BMS}. These massive pion solutions are closer to
the structure of real nuclei. They are more compact, and clustering 
can be observed. For example, the $B=8$ Skyrmion consists of 
two $B=4$ Skyrmions, as in the $\alpha$-particle model of nuclei. The reason 
for this greater compactness is that in the hollow region of the 
shell-like Skyrmions, $U$ is close to $-{\boldsymbol I}$, and 
the pion mass term makes this unstable. 

Many Skyrmions have been found with the help of the rational map ansatz
\cite{HMS}. A rational map is a quotient of two complex polynomials 
$p(z)/q(z)$, and using stereographic projection it can be interpreted 
as a map from the Riemann sphere, $S^2$, to itself. A rational map is an 
exact multi-lump solution of
the $O(3)$ sigma model on $S^2$, this being a Skyrme-type model 
in two dimensions. In three dimensions, the rational 
map is used to encode the angular part of a Skyrmion, and by extending 
it using a radial profile function, one gets useful field 
configurations of the three-dimensional Skyrme model. Part of the Skyrme
energy depends on the coefficients of the rational map and it is
important to optimise the coefficients by minimising the energy, or at
least by getting close to the minimum. After this, a full numerical relaxation 
quickly leads to a shell-like Skyrmion, with the baryon number 
equal to the degree of the map. The approach can be extended to a 
multi-layer ansatz using two or more rational maps \cite{MP}, and
then, with $m=1$, numerical relaxation leads to the more 
compact Skyrmions. 

Rational maps can be constructed with the conjectured symmetry of 
a Skyrmion for a given baryon number. This is helpful, 
although one should check that configurations with different 
symmetries are not of lower energy. The 
coefficients of a rational map are constrained by its symmetry, and
for small degrees only certain symmetries are allowed. For example,
for degree 4 there is an essentially unique rational map with cubic
symmetry, and no map with icosahedral symmetry. This goes some way to
explaining the cubic symmetry of the $B=4$ Skyrmion. However, 
for larger baryon numbers, symmetry does not fix 
the coefficients uniquely. The remaining undetermined coefficients
can be found by numerical optimisation of the 
relevant part of the Skyrme energy \cite{BS3c}, but this 
proves to be very time consuming and ineffective for baryon numbers
beyond about $20$. New techniques to construct near-optimal rational 
maps are presented in this paper, and several new Skyrmion solutions 
have been found using these. They have various baryon numbers up 
to $B=108$, far higher than those of Skyrmions found before. The 
closest comparable solution, previously known, is the cubic 
$B=32$ Skyrmion, which can be obtained using a double rational 
map ansatz \cite{BMS}.

A feature of many Skyrmions with massive pions is that
they look like fields cut out from the infinite Skyrme crystal. We
call these crystal chunks. The crystal with massless pions has a 
primitive cubic structure where each unit cell contains half a unit 
of baryon number, and can be regarded as containing a half-Skyrmion
\cite{KSh,KSh1,CJJVJ}. In neighbouring unit cells the fields repeat 
with an $SU(2)$ twist. 
There is exact periodicity after two lattice spacings, so a true 
cubic unit cell contains eight half-Skyrmions, and hence four units of 
baryon number. In the crystal with massive pions, the half-Skyrmion 
symmetry is slightly broken. The true unit cell remains 
a cube with four units of baryon number. The fields in this unit 
cell are very similar to the isolated $B=4$ Skyrmion. As a 
consequence, many Skyrmion solutions with baryon number a multiple 
of four are crystal chunks. At the same time they look like clusters 
of $B=4$ Skyrmions glued together, analogous to what is expected 
in the $\alpha$-particle model \cite{BMS}.  

The geometrical method used to construct rational maps in
this paper is based on the Skyrme crystal. The zeros and poles of 
the rational map are derived from the locations of half-Skyrmions in 
the crystal \cite{MSkySD}. This requires a conversion of the
Cartesian coordinates in the crystal lattice (relative to a suitable 
origin) to angular coordinates, and then to the Riemann sphere 
coordinate $z$. The simplest rational maps acquire the cubic point 
symmetry of the crystal, but these have restricted values for their 
degrees, resulting in Skyrmions with a restricted set of baryon 
numbers. However, by selecting
subsets of the half-Skyrmions, and by applying a corner-cutting technique that
we will explain below, we are able to construct a large range of 
useful rational maps with lower symmetry. This yields a larger 
set of baryon numbers. We would like to construct Skyrmions with 
every possible baryon number up to $B=300$, as this would potentially 
deal with all nuclei. However, this remains beyond our reach. 

While some of the solutions we find are composed of $B=4$
Skyrmions, there are some exceptions. Two new solutions with $B=20$
have $T_d$ and $D_{2h}$ symmetry, respectively. Both solutions consist 
of four $B=4$ Skyrmions with four $B=1$ Skyrmions between them, rather 
than of five $B=4$ Skyrmions. This resembles the $4\alpha + 4n$ cluster structure recently suggested for $^{20}$O \cite{FKDKO}.

A few solutions presented here were not constructed using a rational
map ansatz. Instead they were found by relaxing an initial
configuration made from a number of $B=4$ Skyrmions glued together 
using the product ansatz \cite{book}. The $B=24$ solutions we have
obtained this way are of lower energy than anything constructed using the
rational map ansatz.

In Section 2 we recall the basic $B=1$ hedgehog Skyrmion,
and the colouring scheme for pion fields that has been found useful 
before, and is used again here. The rational map ansatz and the multi-layer
rational map ansatz are also recalled. In Section 3 we describe the
geometrical method, based on the Skyrme crystal structure, for
constructing many useful rational maps. New Skyrmion solutions
obtained by this method are also discussed, including the Skyrmions 
of highest baryon number that we have found, with $B \ge 100$. 
Section 4 analyses a class of rational maps with cubic or tetrahedral 
symmetry, and some related Skyrmions. Section 5 describes 
the two $B=24$ Skyrmions that we have obtained as clusters of six $B=4$
Skyrmions. Concluding remarks are in Section 6, and our numerical methods are discussed in the Appendix. 

\section{Hedgehogs and rational maps}\news
\subsection{$B=1$ Skyrmion and colouring scheme}

The $B=1$ Skyrmion has an $O(3)$ symmetry (combined
rotations in real space and among the pion fields, together with
an inversion symmetry). It looks like a hedgehog in that the pion 
fields are pointing radially outward from the centre. The ansatz 
for this field configuration is
\be
U({\mathbf x}) = \exp \left( i f(r) \, {\hat{\mathbf x}} 
\cdot {\mathbf{\pauli}}\right) \,,
\label{hedgehog}
\ee
where $r=|{\mathbf x}|$ and ${\hat{\mathbf x}} = {\mathbf x} / r$. This results in $\sigma = \cos f(r)$, $\bpi = \sin f(r) \, \hat{\bf x}$, and
to get a $B=1$ configuration, the boundary conditions
\be
f(0) = \pi \,, \quad f(\infty) = 0
\ee
have to be imposed. Optimising the radial profile function $f(r)$, to
minimise the energy, gives the Skyrmion. Table 1 lists the energies
and other properties of this and other Skyrmions that we have found.

To visualise Skyrmions, a surface of constant baryon density is plotted,
coloured using P.O. Runge's colour sphere. The colours 
indicate the value of the normalised pion field $\hat\bpi = \bpi/|\bpi|$. 
No attempt at colouring is made at points where $\sigma = \pm 1$ and
$\bpi=0$, but these are absent from the surfaces we show. 
The equator of the colour sphere corresponds to $\hat\bpi_3 =
0$. Here, the primary colours red, green and blue show where the field
$\hat\bpi_1+i\hat\bpi_2$ takes the values $1, \exp(i 2\pi /3)$ and $\exp(i
4\pi /3)$, and the intermediate colours yellow, cyan and magenta show
the values $\exp(i \pi /3), -1$ and $\exp(i 5\pi /3)$. The
$\hat\bpi_3$-value is assigned to the ``lightness'' colour attribute 
so that white and black, at the poles on the colour sphere,
show where $ \hat \bpi _3 = \pm 1$, respectively.

\begin{figure}[ht]
\centering
\includegraphics[width=7cm]{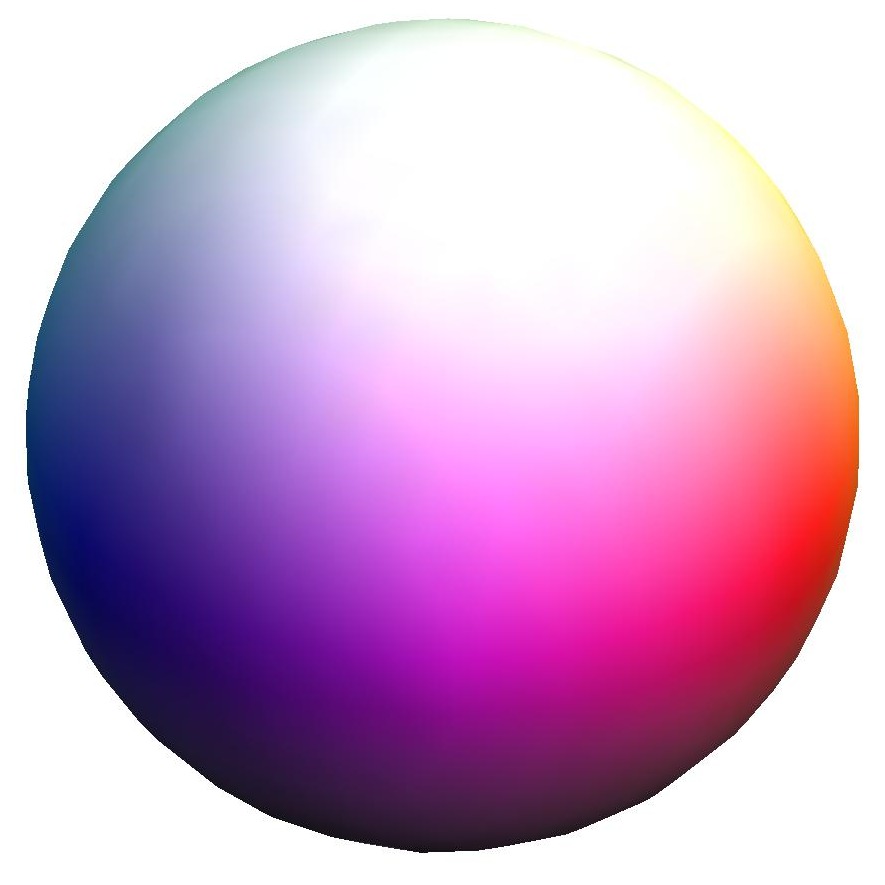}
\caption{Colour sphere, and $B=1$ Skyrmion}
\label{fig:B_1}
\end{figure}

The hedgehog form of the $B=1$ Skyrmion means that on a spherical
surface, the colouring reproduces the colour sphere itself (see Figure
\ref{fig:B_1}).

Because the pion fields are scalar fields, charges (sources) of equal 
sign attract. As a consequence, parts of Skyrmions with the same colour
tend to attract, and low energy configurations can be constructed by
gluing Skyrmions together with colours matching. This is what makes 
the colouring so useful. However for $B>1$ there is some frustration, that is, non-matching colours, 
as one can see from some of the figures in this paper. If 
there were no frustration, as occurs for example with a $B=1$ and 
$B=-1$ Skyrmion pair, then the Skyrmions could annihilate. 

\subsection{Single rational map ansatz}

The rational map ansatz \cite{HMS} generates approximate Skyrmions with
separated radial and angular dependence, generalising the hedgehog 
ansatz (\ref{hedgehog}). The radial part is again given by a profile function 
$f(r)$ satisfying the boundary conditions $f(0) = \pi$ and 
$f(\infty) = 0$. It is assumed that $f(r)$ decreases 
monotonically as $r$ increases. The angular part is determined by a 
rational function $R(z) = p(z)/q(z)$, where $p(z)$ and $q(z)$ are 
polynomials. Here $z$ is the complex coordinate on the Riemann
sphere defined, via stereographic projection, as
$z= \tan \left( \frac{1}{2}\theta \right) \exp(i \phi)$, where $\theta$ and $\phi$ are 
the usual spherical polar coordinates. The $z$-coordinate is related to the Cartesian
coordinates on the unit sphere $S^2 \subset \bR ^3$ via the formula
\be
z = \frac{{\bf \hat n}_1 + i {\bf \hat n}_2}{1 + {\bf \hat n}_3} \,,
\label{znhat}
\ee
where ${\bf \hat n}$ is the outward normal vector on the unit sphere.
The inverse of this formula is
\be
{\bf \hat n} _z = \frac{1}{1 + |z|^2} 
(z + \bar z \, , \, i(\bar z -z) \, , \, 1 - |z|^2) \,.
\label{nhat}
\ee
The rational map ansatz, combining the rational map $R(z)$
with the radial profile function $f(r)$, is defined by
\bea
U(r,z) &=& \exp(i f(r) \, {\bf \hat n} _{R(z)} \cdot \pauli) \nonumber \\
&=& \cos f(r) \, {\boldsymbol I} + i \sin f(r) \, {\bf \hat n} _{R(z)} 
\cdot \pauli \,,
\label{ansatz}
\eea
where ${\bf \hat n} _{R(z)}$ is defined analogously to (\ref{nhat}).
This generalises the hedgehog ansatz (\ref{hedgehog}), which can be 
recovered by setting $R(z) = z$.

\begin{figure}[ht]
\centering
\includegraphics[width=5cm]{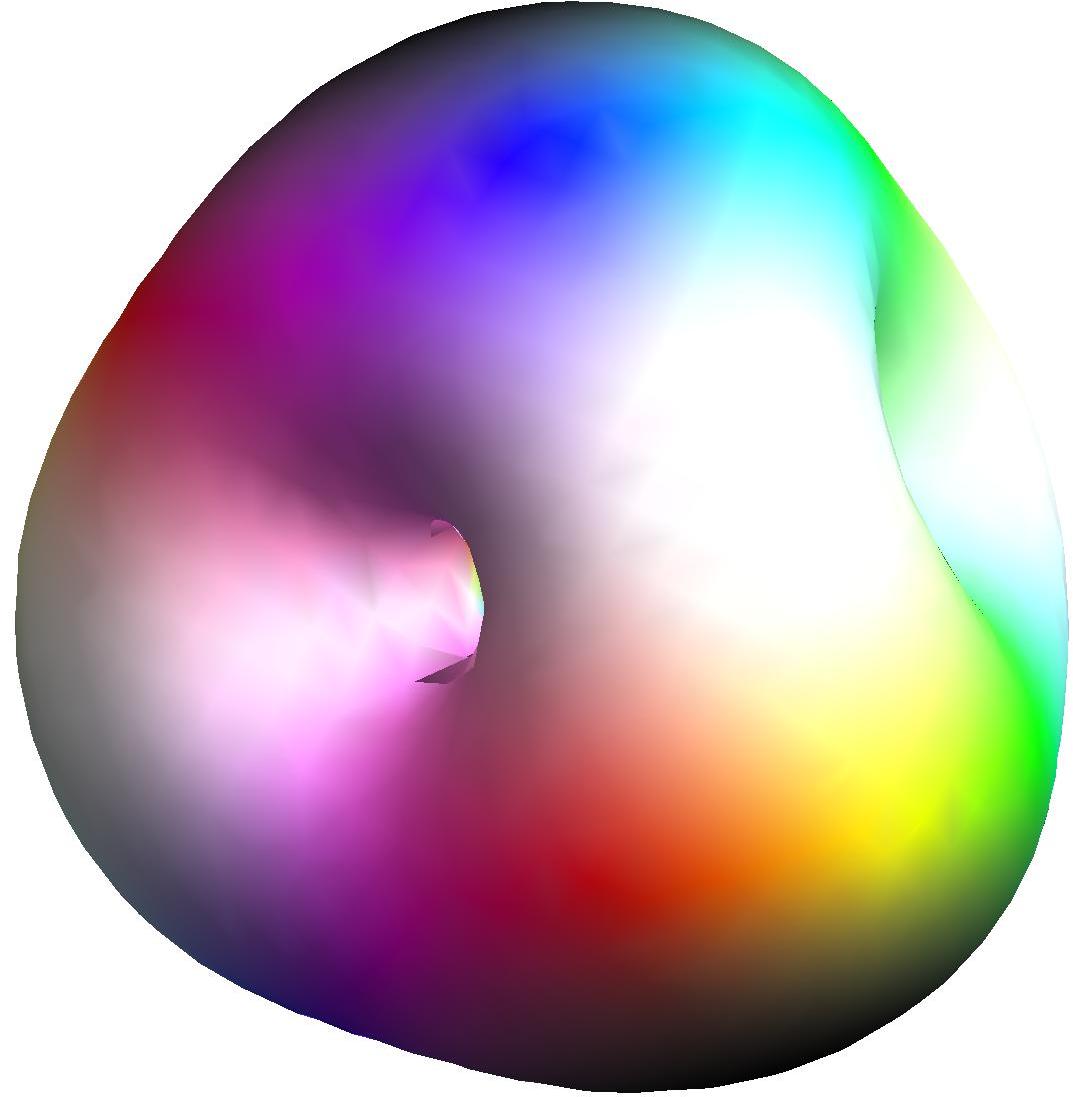}
\hspace{1cm}
\includegraphics[width=5cm]{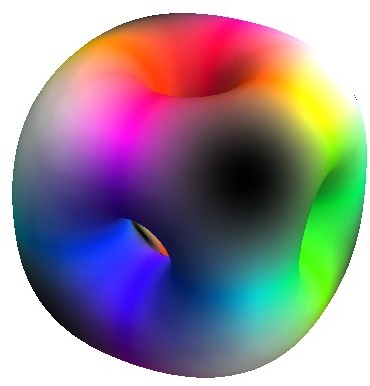}
\caption{$B=3$ tetrahedral (left) and $B=4$ cubic (right) Skyrmions}
\label{fig:B_4}
\end{figure}

The baryon number $B$ is the topological degree of the rational map 
$R: S^2 \rightarrow S^2$, which is also the algebraic degree of
$R(z)$, the higher of the degrees of the polynomials $p(z)$ and
$q(z)$. Below are some of the well known rational maps of high
symmetry which approximate the Skyrmions with baryon numbers 
$B = 1, 2, 3, 4$:
\begin{align}
R(z) \ &= \ z \, , & \qquad R(z) \ & = \ z^2\, , \nonumber \\
R(z) \ &= \ \frac{z^3 - \sqrt{3} i z}{\sqrt{3} i z^2 - 1} \, , & \qquad R(z) \ & = \ \frac{z^4 + 2\sqrt{3} i z^2 + 1}{z^4 - 2 \sqrt{3} i z^2 + 1} \, .
\label{eq:lowbratmaps}
\end{align}
Optimising the profile function $f(r)$ gives approximate Skyrmions,
but true Skyrmions can only be found by further numerical relaxation.
The true Skyrmions have the same symmetries as their rational maps,
but for $B>1$ their angular dependence does vary with radius, contrary 
to what the rational map ansatz allows. The symmetry groups
are $O(3), \ D_{\infty h}, \ T_d, \ O_h$ for $B=1,2,3,4$
respectively. The $B=3$ and $B=4$ Skyrmions are shown in Figure \ref{fig:B_4}.

\subsection{Multi-layer rational map ansatz}

The single rational map ansatz works well for Skyrmions of small
baryon number, but for larger Skyrmions, a multi-layered structure is
needed. The double rational map ansatz was the first extension to 
be considered \cite{MP}. It uses two rational maps denoted
$R^\mathrm{in}(z)$ and $R^\mathrm{out}(z)$. A monotonic profile 
function $f(r)$ is needed, taking an extended range of values 
with $f(0) = 2 \pi$ and $f(\infty) = 0$. Let $r_1$ be the radius where
$f(r_1) = \pi$. The ansatz is the same as (\ref{ansatz}), but 
$R(z) = R^\mathrm{in}(z)$ for $0 \le r \le r_1$ and $R(z) =
R^\mathrm{out}(z)$ for $r_1 \le r < \infty$. On the sphere of radius $r_1$, 
the Skyrme field is $U = - \boldsymbol I$. The total baryon number is 
the sum of the degrees of the maps $R^\mathrm{in}$ and $R^\mathrm{out}$.  

In the $K$-layer ansatz, the profile
function needs to take the values $f(0) = K \pi$, $f(\infty) = 0$ with
the $k$-th rational map $R^k (z)$ used in the region $r_{k-1} \le r \le r_k$
for $k = 1,2, \cdots, K$. Here $f(r_k) = (K-k) \pi$, with $r_0 = 0 \,, 
r_K = \infty$. The total baryon number is the sum of the degrees of
all the rational maps.

It is best if the maps in the different layers share a 
substantial amount of symmetry. The maps in neighbouring layers 
need to be carefully oriented so that colours match as far as 
possible. This requirement fixes certain relative signs that are 
not determined by symmetry alone. The multi-layer ansatz is useful, 
but not so close to Skyrmion solutions as the single layer version. 
After relaxation, the solutions do not have $U = \pm \boldsymbol I$ 
on complete spheres, but instead at isolated points.
 
\section{Geometrical construction of rational maps}\news
\subsection{Skyrmions from the Skyrme crystal}

Rational maps of low degree can be constructed using symmetry algebra 
to fix the coefficients of the polynomials $p(z)$ and $q(z)$. 
Generally, however, symmetry leaves some coefficients undetermined,
and these have in the past been found
numerically, by minimising an expression that arises when 
the rational map ansatz is inserted into the Skyrme energy
\cite{HMS}. For degrees beyond about 20, however, this minimisation 
become intractible, and a new approach is needed.

The Skyrme crystal, for massless pions, is made of half-Skyrmions \cite{KSh,KSh1,CJJVJ}. The
field values at lattice points of the crystal are known precisely,
because of the crystal symmetry constraints. In a convenient
orientation, the $\hat\bpi_3$ field takes the values $\pm 1$ at 
alternating lattice points (white and black in our colour scheme). 
These values correspond to the zeros and poles of a rational map 
$R(z)$, that is, to the roots of the polynomials $p(z)$ and 
$q(z)$.

This observation \cite{MSkySD} is the basis for a geometrical 
construction of a range of useful rational maps, which lead to new Skyrmions.
In particular, the construction leads to the cubic Skyrmions with baryon number 
$B = 4 n^3$, with $n \in \mathbb{N} ^{+}$, resembling chunks of the 
Skyrme crystal. The first construction of the $B=32$ Skyrmion (the
$n=2$ example) was by manipulating the field of the Skyrme crystal 
directly \cite{Ba}. However, it is easier to use the intermediary 
of an $n$-layer rational map ansatz. The real advantage of the
rational map approach is that one can fairly easily vary the degrees 
of the maps and generate Skyrmions with baryon numbers not of the 
form $4 n^3$. 

\begin{figure}[ht]
\centering
\hspace{4cm}
\includegraphics[width=9cm]{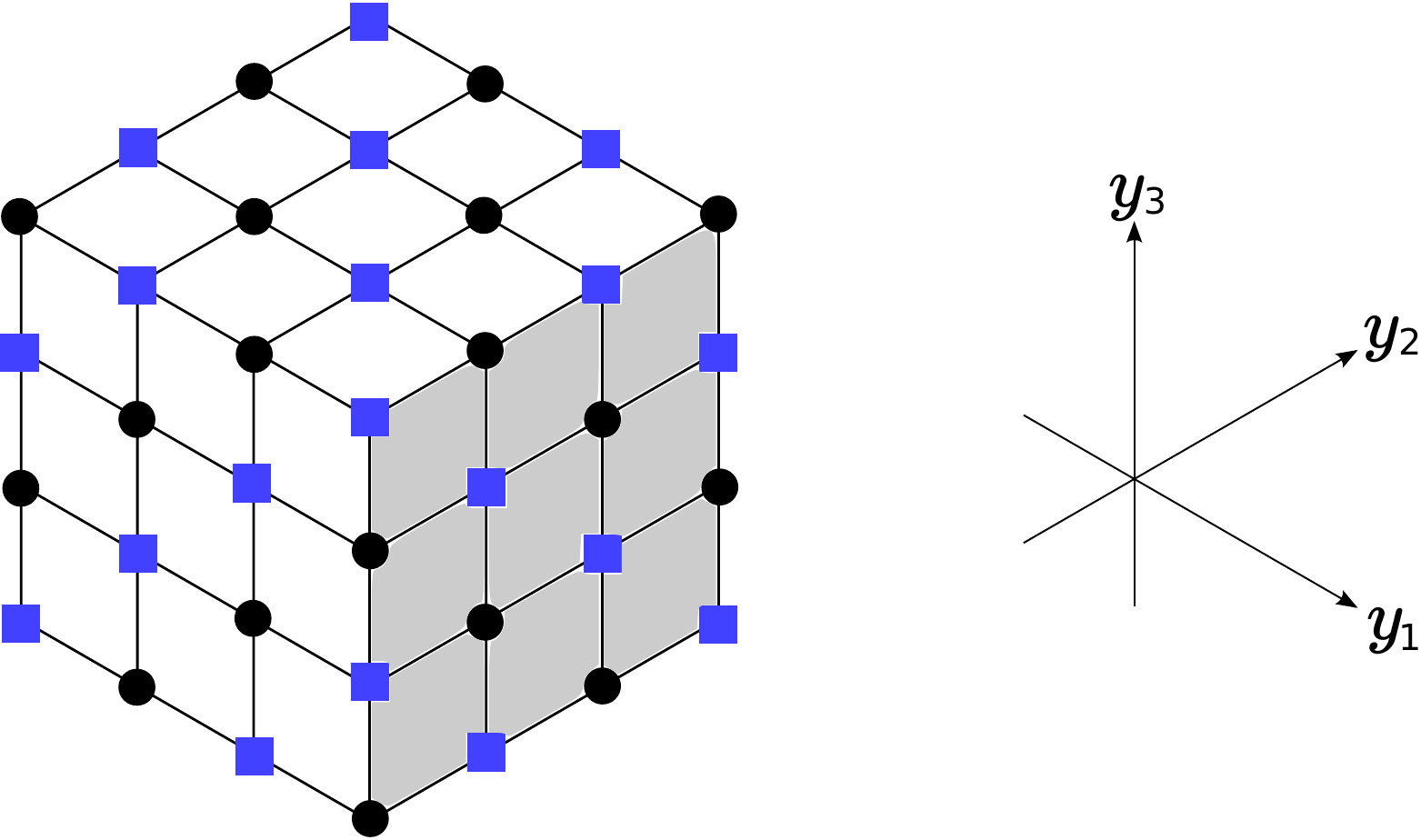}
\caption{$4 \times 4 \times 4$ Cubic grid}
\label{fig:cube_grid}
\end{figure}

Here is how we rederived the $B=32$ Skyrmion. Figure 
\ref{fig:cube_grid} shows the $4 \times 4 \times 4$ grid of half-Skyrmion
locations that occur in a cubic chunk of the crystal. The (black)
circles and (blue) squares are used as the zeros and poles,
respectively, for the construction of the rational maps. The grid needs to 
be separated into two layers. The outer layer has 56 points and the 
inner (hidden) layer has $2 \times 2 \times 2 = 8$ points. The corresponding 
rational maps have degrees 28 and 4.

We set up scaled Cartesian coordinates $(y_1,y_2,y_3)$, with the origin 
at the centre of the chunk, and lattice points having half-integer 
coordinates. The distance to the origin is denoted by $\rho$. For example, 
the inner layer has its eight points at $(\pm \half, \pm \half, \pm \half)$, 
with $\rho=\sqrt{3}/2$. The outer corner points are at 
$(\pm 3/2, \pm 3/2, \pm 3/2)$, with $\rho=3\sqrt{3}/2$. Other points 
in the outer layer are at distances $\rho=\sqrt{19}/2$ (points on edges) and $\rho=\sqrt{11}/2$ (face points).

The Riemann sphere coordinate for any of these points is
\be
z = \frac{y_1 + i y_2}{\rho + y_3} \,,
\ee
a variant of (\ref{znhat}). From the inner layer of points, we 
construct a degree 4 map. The numerator $p(z)$ has roots at
$z = \pm \frac{1-i}{\sqrt{3} + 1}$, $\pm \frac{1+i}{\sqrt{3} - 1}$ 
and the denominator $q(z)$ has roots at $z = \pm \frac{1+i}{\sqrt{3} + 1}$, 
$\pm \frac{1-i}{\sqrt{3} -  1}$. This gives
\begin{align}
R(z) &= \frac{(z + \frac{1-i}{\sqrt{3} + 1})(z - \frac{1-i}{\sqrt{3} + 1})(z + \frac{1+i}{\sqrt{3} - 1})(z - \frac{1+i}{\sqrt{3} - 1})}{(z + \frac{1+i}{\sqrt{3} + 1})(z - \frac{1+i}{\sqrt{3} + 1})(z + \frac{1-i}{\sqrt{3} - 1})(z - \frac{1-i}{\sqrt{3} - 1})} \nonumber \\ 
&= \frac{z^4 + 2\sqrt{3} i z^2 + 1}{z^4 - 2\sqrt{3} i z^2 + 1} \,,
\label{deg4}
\end{align}
the degree 4 rational map with cubic
symmetry, related to the $B=4$ Skyrmion. Cubic symmetry requires 
that the overall coefficient is of magnitude 1, and here it is set to 1. 

A similar procedure gives the degree 28 rational map of the outer
layer. The numerator and denominator are expressed as products of
their linear factors. The map is
\begin{align}
R(z) =& \frac{(z - \frac{3-i}{\sqrt{19}+3})(z - \frac{1-3i}{\sqrt{19}+3})(z - \frac{-3+i}{\sqrt{19}+3})(z - \frac{-1+3i}{\sqrt{19}+3})(z - \frac{3+3i}{\sqrt{19}-1})(z - \frac{-3+3i}{\sqrt{19}+1})}{(z - \frac{3+i}{\sqrt{19}+3})(z - \frac{1+3i}{\sqrt{19}+3})(z - \frac{-3-i}{\sqrt{19}+3})(z - \frac{-1-3i}{\sqrt{19}+3})(z - \frac{3+3i}{\sqrt{19}+1})(z - \frac{-3+3i}{\sqrt{19}-1})} \nonumber \\
\times & \frac{(z - \frac{-3-3i}{\sqrt{19}-1})(z - \frac{3-3i}{\sqrt{19}+1})(z - \frac{3+i}{\sqrt{19}-3})(z - \frac{-1-3i}{\sqrt{19}-3})(z - \frac{-3-i}{\sqrt{19}-3})(z - \frac{1+3i}{\sqrt{19}-3})}{(z - \frac{-3-3i}{\sqrt{19}+1})(z - \frac{3-3i}{\sqrt{19}-1})(z - \frac{3-i}{\sqrt{19}-3})(z - \frac{-1+3i}{\sqrt{19}-3})(z - \frac{-3+i}{\sqrt{19}-3})(z - \frac{1-3i}{\sqrt{19}-3})} \nonumber \\
\times & \frac{(z - \frac{1+i}{\sqrt{11}+3})(z - \frac{-1-i}{\sqrt{11}+3})(z - \frac{3+i}{\sqrt{11}+1})(z - \frac{3-i}{\sqrt{11}-1})(z - \frac{1+3i}{\sqrt{11}+1})(z - \frac{-1+3i}{\sqrt{11}-1})}{(z - \frac{1-i}{\sqrt{11}+3})(z - \frac{-1+i}{\sqrt{11}+3})(z - \frac{3+i}{\sqrt{11}-1})(z - \frac{3-i}{\sqrt{11}+1})(z - \frac{1+3i}{\sqrt{11}-1})(z - \frac{-1+3i}{\sqrt{11}+1})} \nonumber \\
\times & \frac{(z - \frac{-3-i}{\sqrt{11}+1})(z - \frac{-3+i}{\sqrt{11}-1})(z - \frac{-1-3i}{\sqrt{11}+1})(z - \frac{1-3i}{\sqrt{11}-1})(z - \frac{-1+i}{\sqrt{11}-3})(z - \frac{1-i}{\sqrt{11}-3})}{(z - \frac{-3-i}{\sqrt{11}-1})(z - \frac{-3+i}{\sqrt{11}+1})(z - \frac{-1-3i}{\sqrt{11}-1})(z - \frac{1-3i}{\sqrt{11}+1})(z - \frac{1+i}{\sqrt{11}-3})(z - \frac{-1-i}{\sqrt{11}-3})} \nonumber \\
\times & \frac{(z - \frac{-1-i}{\sqrt{3}+1})(z - \frac{1+i}{\sqrt{3}+1})(z - \frac{-1+i}{\sqrt{3}-1})(z - \frac{1-i}{\sqrt{3}-1})}{(z - \frac{1-i}{\sqrt{3}+1})(z - \frac{-1+i}{\sqrt{3}+1})(z - \frac{1+i}{\sqrt{3}-1})(z - \frac{-1-i}{\sqrt{3}-1})} \text{ .}
\label{deg28}
\end{align}
Cubic symmetry and matching to the inner layer map fixes the overall 
coefficient to be 1. The linear factors could be multiplied out,
giving a map of the structure discussed in \cite{BMS}. However, there is 
little reason to do this: the linear factor representation makes it easier 
to check symmetry and to recall the coordinates of the half-Skyrmions. 
It also avoids problems with overflowing numerics.

This degree 28 rational map is not optimal energetically, but it is
close to optimal because the zeros and poles are approximately evenly spread
over the Riemann sphere, and they are rather well separated from each other.
Using these maps of degrees 4 and 28 in the double rational map ansatz
as initial data, with a simple, piecewise-linear radial profile function,
we successfully recover the cubically symmetric $B=32$ Skyrmion using 
the numerical relaxation algorithm developed in \cite{Fei}.

\begin{figure}[ht]
\centering
\includegraphics[width=12cm]{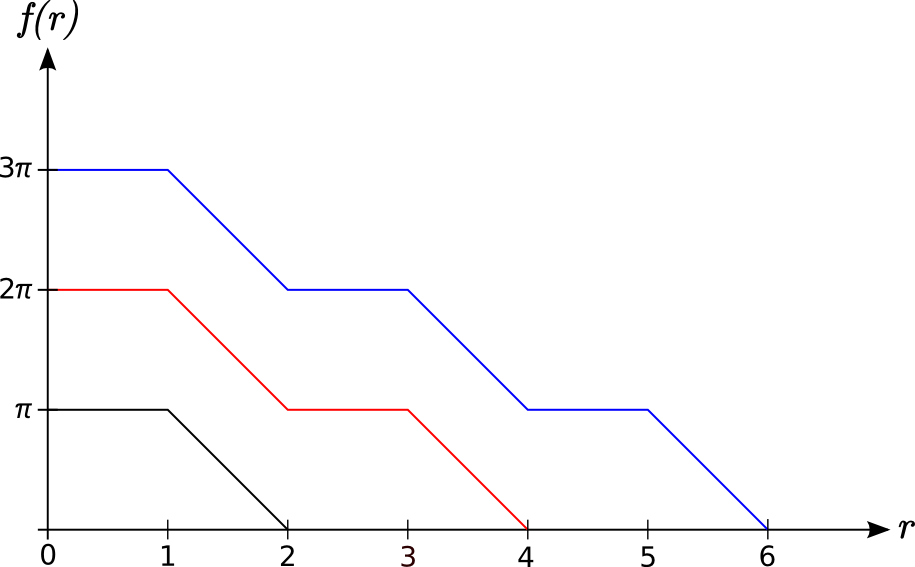}
\caption{Profile functions $f(r)$ used for initial data}
\label{fig:profile_function}
\end{figure}

The profile functions we have used are shown in Figure 
\ref{fig:profile_function}. The profile functions coloured black, red
and blue are used in the single, double and triple rational map ansatz,
respectively. Since $f(r)$ is an integer multiple of
$\pi$ in finite intervals of $r$, the Skyrme field $U$ initially takes the 
value $\pm \boldsymbol I$ in spherical shells of finite thickness.

\subsection{Exploring $B=24$ -- $31$ solutions using the corner cutting method}

The idea of cutting single Skyrmions from the corners of the $B=32$ Skyrmion was proposed in \cite{MSkySD} to generate new solutions 
with baryon numbers from $B=31$ down to $B=24$. The corner cutting
procedure is done on the degree 28 rational map (\ref{deg28}); 
it is best understood using the geometrical approach to rational 
maps presented above. The inner degree 4 map is unchanged.

The simplest way to remove a Skyrmion from a rational map is to decrease its degree by one by merging a zero and a pole.
Linear factors cancel in the numerator and denominator. Say $R(z)$ has a zero at 
$a$ and a pole at $b$. Then eliminating one Skyrmion is done by 
setting $a=b$:
\be
R(z)\bigg|_{a=b}= \frac{p(z)}{q(z)} \bigg|_{a=b} 
= \frac{(z-a)P(z)}{(z-b)Q(z)}\bigg|_{a=b} = \frac{P(z)}{Q(z)} \,,
\ee
where $P(z)$ and $Q(z)$ are the remaining parts of the polynomials 
$p(z)$ and $q(z)$.

\begin{figure}[ht]
\centering
\includegraphics[width=12cm]{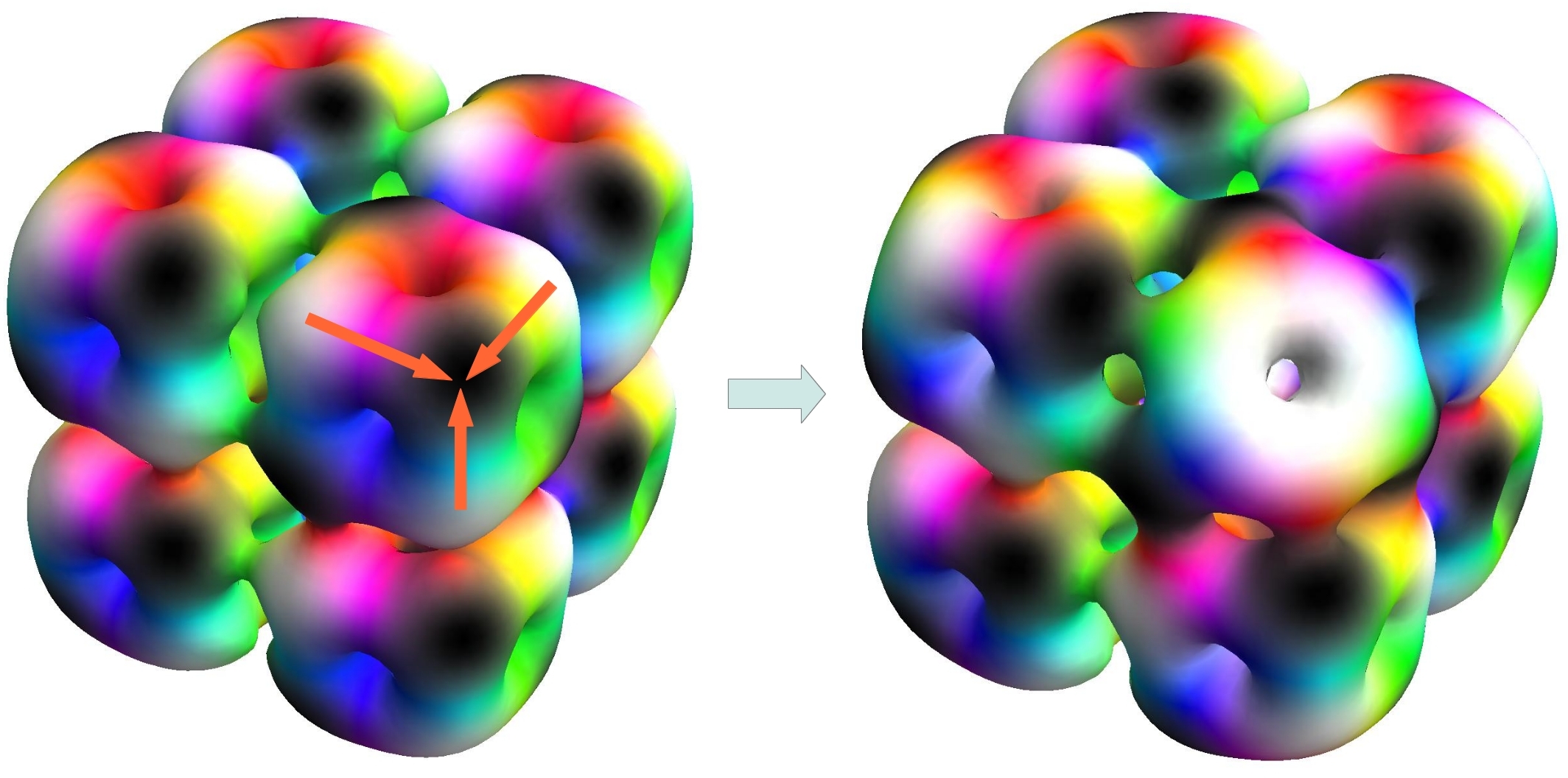}
\caption{Moving three zeros (white) towards a pole (black), to
 produce the $B=31$ Skyrmion}
\label{fig:Cut_corner}
\end{figure}

We could use this method on one of the corner poles (or zeros) of the
map (\ref{deg28}) and one of its neighbouring zeros (or 
poles), but this destroys all the symmetry, which is not desirable. Rather, 
to preserve as much symmetry as possible, the three neighbouring zeros 
are moved simultaneously towards a corner pole. The pole cancels against one
of the zeros, leaving a double zero at the corner after cancellation 
(see Figure \ref{fig:Cut_corner}). Notice that what had been a black
corner becomes a white one with a hole. The $O_h$ symmetry is broken down to 
$C_{3v}$. After numerical relaxation, this method gives a new 
stable Skyrmion with $B=31$. This corner-cutting procedure can be 
repeated up to eight times. At each corner, three poles are merged 
with one zero, or three zeros with one pole, and the result is either 
a double pole or a double zero. These persist in the Skyrmion
solutions as holes in the baryon density at corners that have been cut.

\begin{figure}[h!]
        \begin{subfigure}[b]{0.3\textwidth}
                \centering
                \includegraphics[width=\textwidth]{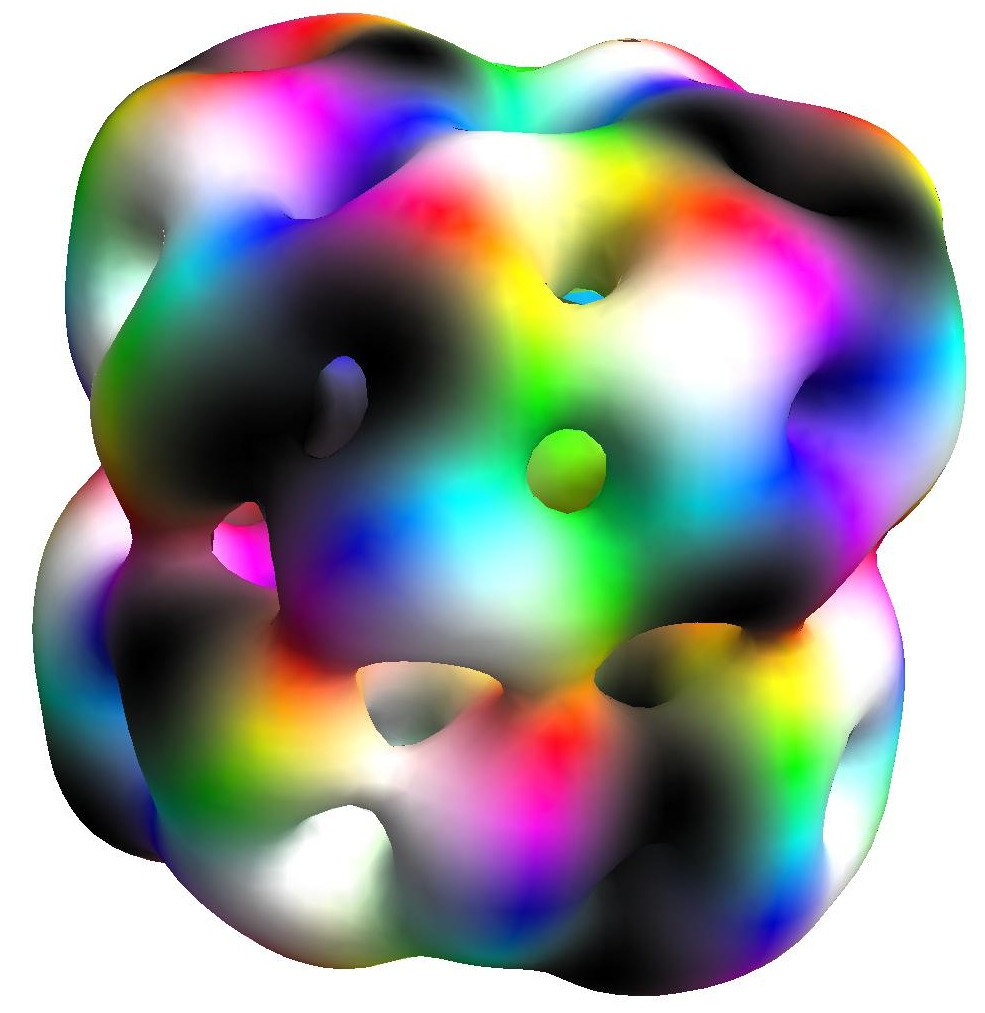}
                \caption{$B=26$}
                \label{fig:B_26}
        \end{subfigure}
        ~ 
        \begin{subfigure}[b]{0.3\textwidth}
                \centering
                \includegraphics[width=\textwidth]{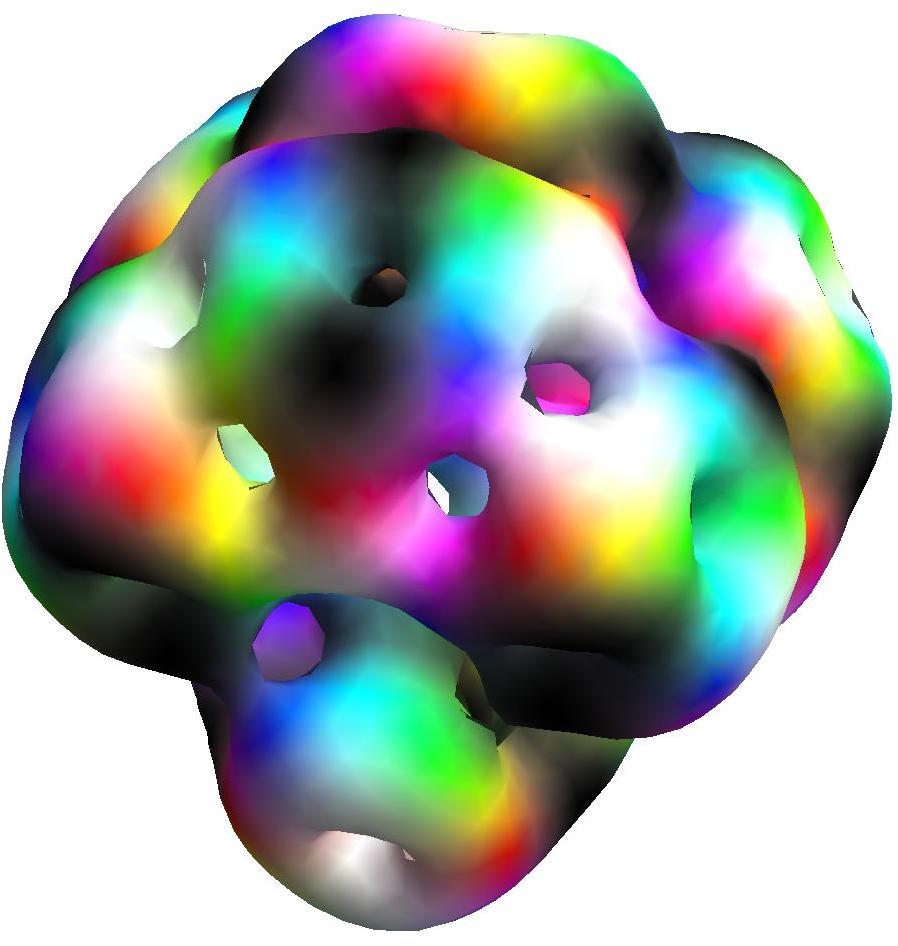}
                \caption{$B=25$}
                \label{fig:B_25}
        \end{subfigure}
        ~ 
        \begin{subfigure}[b]{0.3\textwidth}
                \centering
                \includegraphics[width=\textwidth]{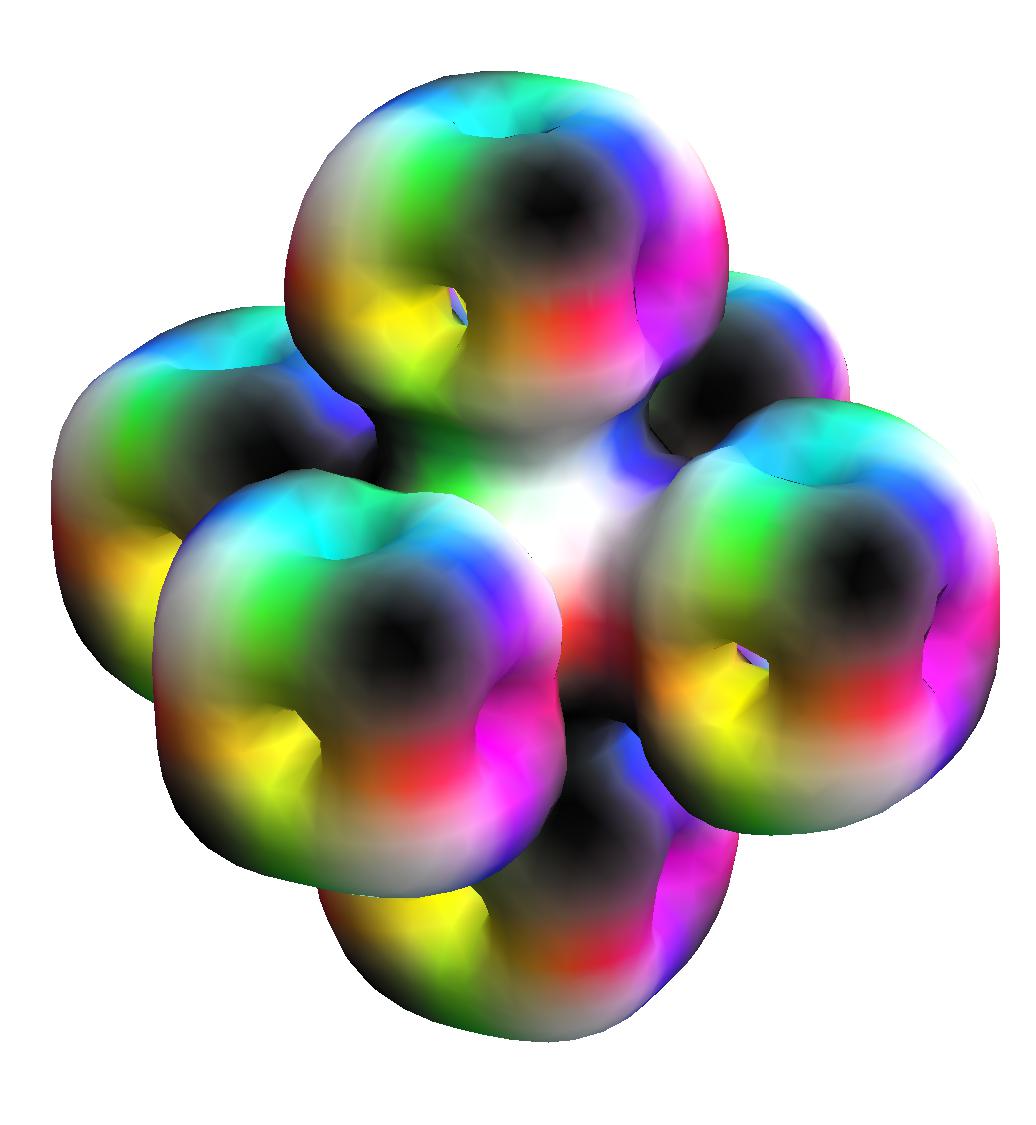}
                \caption{$B=24$}
                \label{fig:B_24}
        \end{subfigure}
\caption{Skyrmions from cutting corners}\label{fig:B_32CC}
\end{figure}

We have obtained solutions from $B=24$ to $B=31$ using this method. 
The $B=32$ Skyrmion looks like eight copies of the cubic $B=4$ Skyrmion. 
In the $B=31$ Skyrmion, one of the $B=4$ cubes becomes a slightly deformed
$B=3$ tetrahedral Skyrmion. The remaining seven hardly change, because the 
interactions between Skyrmions at the corners are weak. When this corner 
cutting is repeated, further $B=4$ cubes are replaced by $B=3$
tetrahedra. In order to preserve as much symmetry as possible, we cut
the corners as follows. For $B=30$, a pair of
diagonally opposite corners are cut and $D_{3d}$ symmetry
remains. For $B=29$, we need to remove three corners and cutting
diagonally opposite corners is no longer a desirable option. Instead,
three corners, with each pair face-diagonally opposite, are cut. 
For $B=28$, four corners forming a tetrahedron are removed and $T_d$
symmetry remains. One further corner is removed to
generate the $B=27$ Skyrmion. For $B=26$, just two diagonally opposite
corners are left uncut. The field relaxation is initially
similar to the higher $B$ cases, but now there is a change. The anticipated 
final $D_{3d}$-symmetric structure is not stable. The six $B=3$ 
Skyrmions at the cut corners move towards the plane mid-way between 
the uncut corners. Then the $B=4$ cubes move in opposite ways and 
join up, one each, with three of the six $B=3$ Skyrmions forming a 
structure consisting of two similar $B=13$ clusters 
(see Figure \ref{fig:B_26}). The $B=13$
cluster has only one reflection symmetry and smaller Skyrmion 
clusters could not be identified clearly within it.

For $B=25$, seven corners are cut initially. However,
the relaxed $B=25$ Skyrmion again does not have the shape expected from corner cutting (see Figure \ref{fig:B_25}). The relaxed Skyrmion only has a $C_{1h}$ reflection symmetry, but we can still vaguely identify cubic $B=4$ Skyrmions and tetrahedral $B=3$ Skyrmions within the cluster.

$B=24$ is the most interesting case. Cutting all eight corners from
the $B=32$ Skyrmion produces a cubically symmetric solution that is
best thought of as six $B=4$ cubes at the vertices of the dual 
octahedron of the $B=32$ cube, rather than eight $B=3$ Skyrmions at the vertices of the 
original cube. Some half-Skyrmions of the $B=4$ 
cubes are acquired from the $B=3$ Skyrmions (see Figure
\ref{fig:B_24}, and note the fate of the three black regions 
near a white corner).

\begin{figure}[ht]
\centering
\includegraphics[trim = 0mm 30mm 0mm 15mm, clip, width=\textwidth]{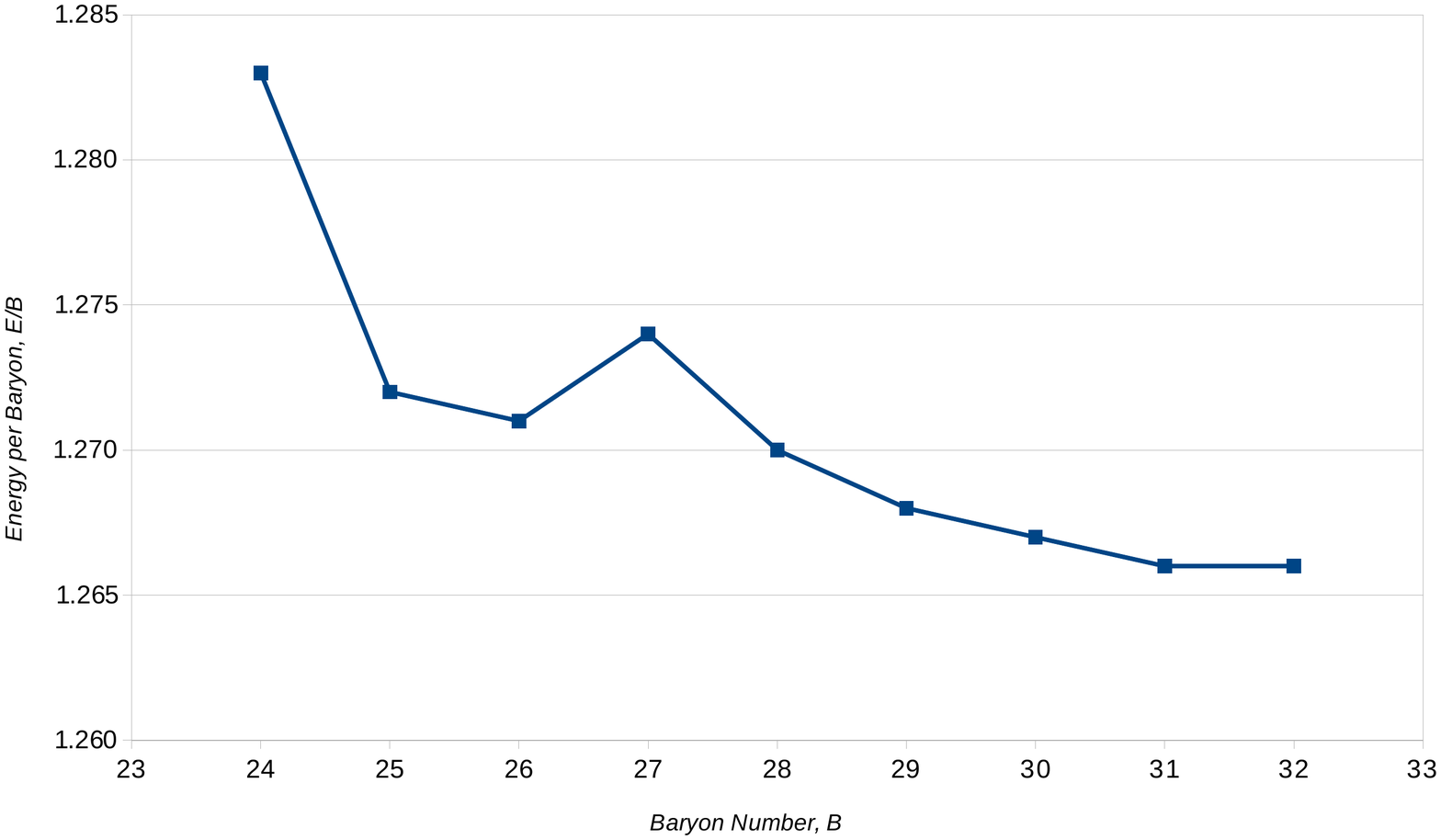}
\caption{Energy per baryon for $B=24$ to $B=32$}
\label{fig:E_per_B}
\end{figure}

Figure \ref{fig:E_per_B} shows the energy per baryon, $E/B$, of the 
Skyrmions found using this corner cutting technique. (They are also
tabulated in Table 1.) For $B=27$ to $B=31$, the structures of the 
Skyrmions are pretty close to that of the parent $B=32$ solution. 
The energy per baryon varies smoothly in this range. 
For solutions from $B=24$ to $B=26$, however, the energy is lower
than the extrapolation of this smooth behaviour. This shows that 
corner cutting is only a first step and further relaxation occurs 
due to the interaction between corners.

\subsection{$B=100$ -- $108$ solutions}

\begin{figure}[h!]
\centering
        \begin{subfigure}[b]{0.3\textwidth}
                \centering
                \includegraphics[width=\textwidth]{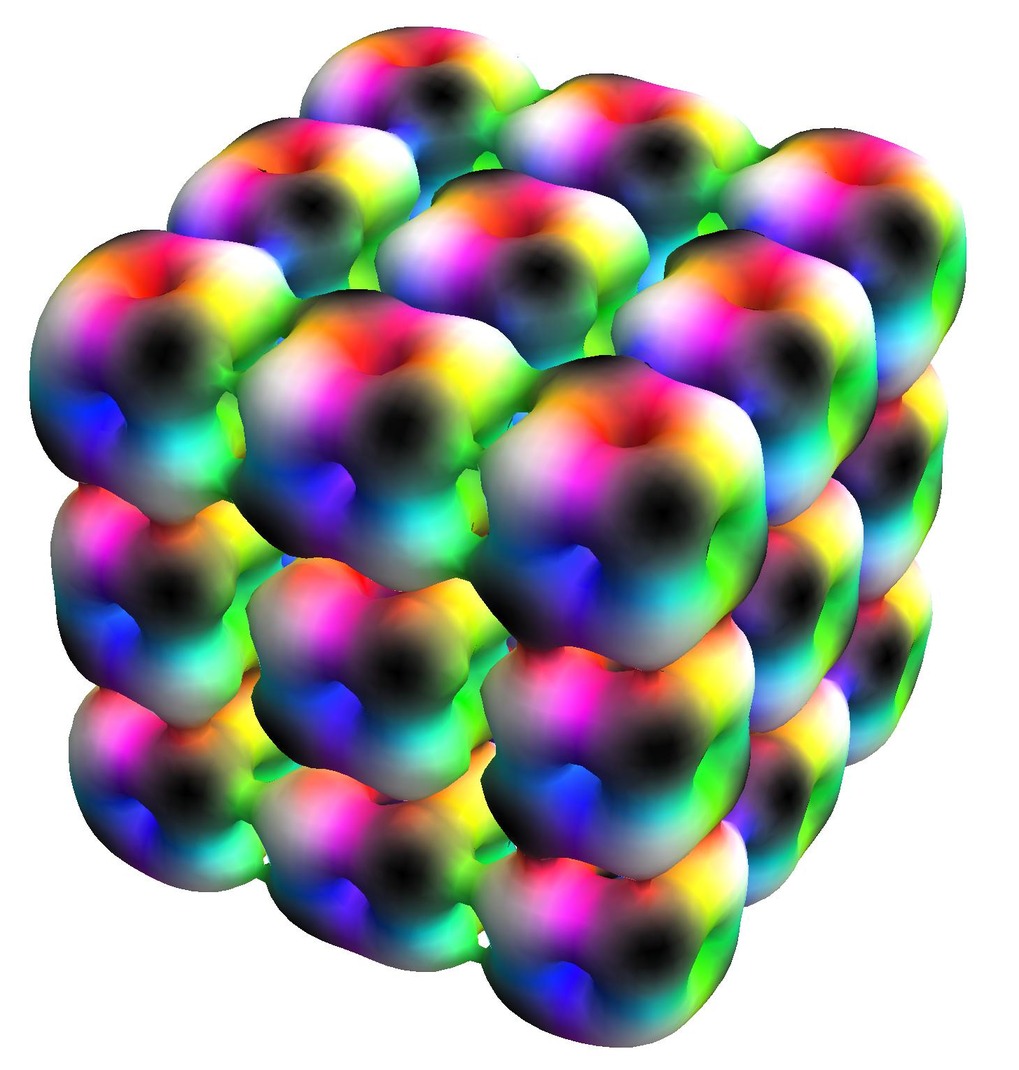}
                \caption{$B=108$ Skyrmion \newline}
                \label{fig:B_108}
        \end{subfigure}
 	\quad 
        \begin{subfigure}[b]{0.3\textwidth}
                \centering
                \includegraphics[width=\textwidth]{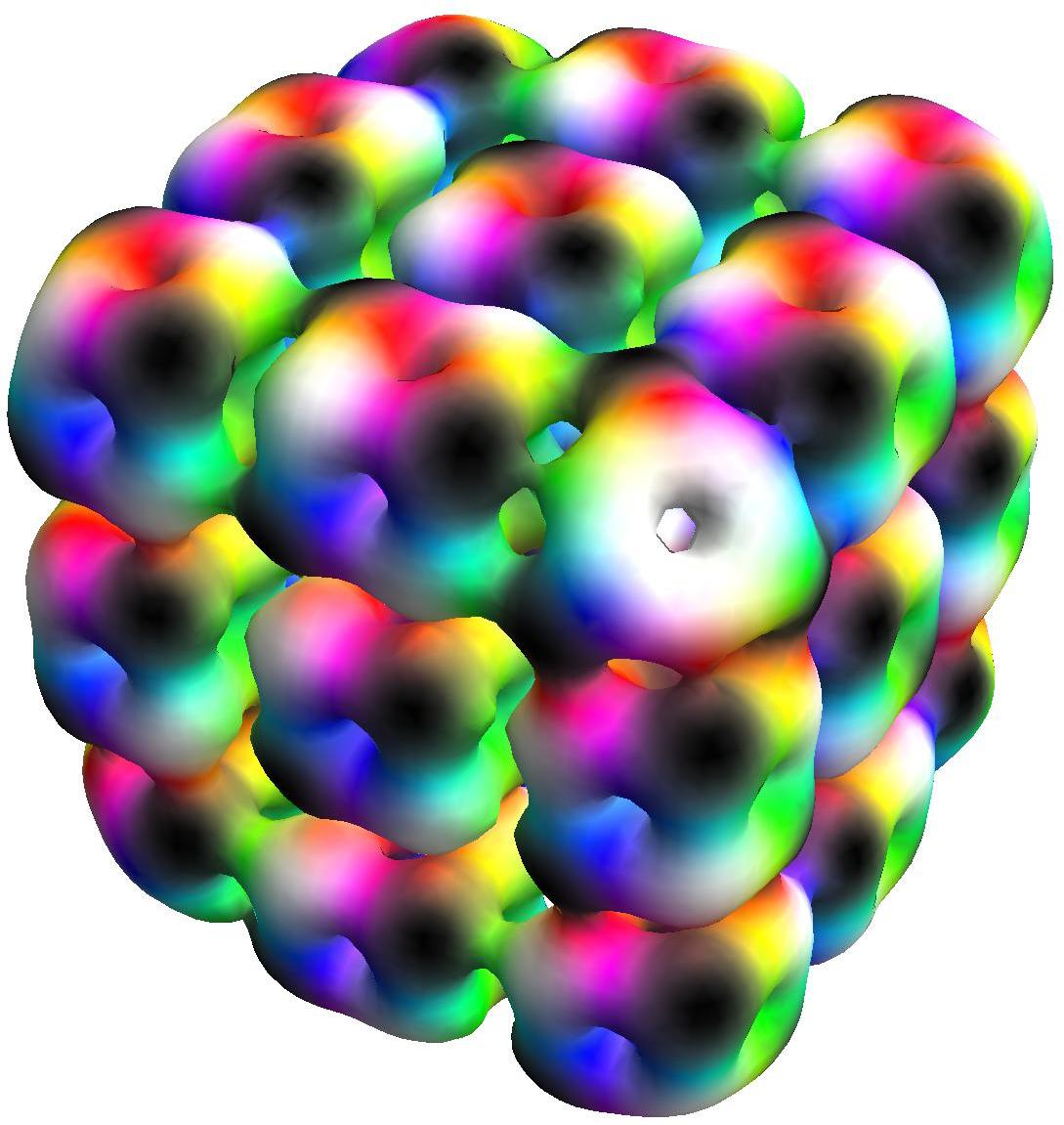}
                \caption{$B=104$ Skyrmion with 4 corners cut}
                \label{fig:B_104}
        \end{subfigure}%
        \quad ~ 
        \begin{subfigure}[b]{0.3\textwidth}
                \centering
                \includegraphics[width=\textwidth]{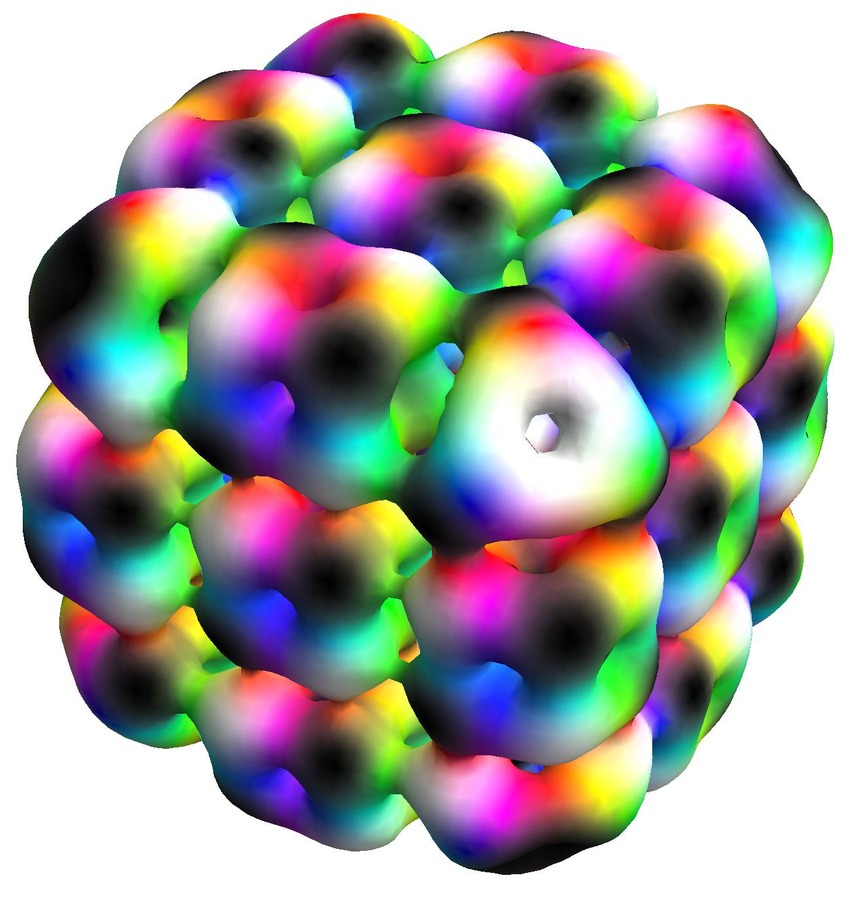}
                \caption{$B=100$ Skyrmion with 8 corners cut}
                \label{fig:B_100}
        \end{subfigure}
        \caption{Skyrmions from triple-layer rational maps}
        \label{fig:Triple}
\end{figure}

The next in the sequence of $B=4n^3$ cubic Skyrmions is $B=108$.
Using a $6 \times 6 \times 6$ cubic grid, we have constructed a 
degree $76$ outer rational map (analogous to (\ref{deg28}) but too long to 
write out). This is 
combined with the degree $28$ and degree $4$ maps from the $B=32$ 
Skyrmion, as middle and inner maps. A suitable profile function 
$f(r)$ running from $3\pi$ to $0$ is used for the initial configuration (see Figure \ref{fig:profile_function}). 
After relaxation, the stable Skyrmion shown in Figure \ref{fig:B_108} was
obtained. It has the familiar structure of touching $B=4$ subunits
all with the same orientation.

The next step is to remove single Skyrmions from the corners. 
This was done, as in the $B=32$ case, by merging three zeros (poles) 
of the outer map with a pole (zero) at each corner. This 
procedure generates Skyrmions with all baryon numbers from $107$ 
down to $100$. The cubic structure is locally retained, except 
that the $B=4$ cubes at the corners are replaced by $B=3$
tetrahedra. Recall that for the $B=32$ Skyrmion there was little 
structural change observed until six or more corner Skyrmions were 
removed. We find the $B=108$ Skyrmion to be more stable to corner
cutting as the corners are further away from each other. Removing
first four, and then all eight corners gives the 
tetrahedral $B=104$ Skyrmion and cubic $B=100$ Skyrmion shown in 
Figures \ref{fig:B_104} and \ref{fig:B_100}.

\section{Rational maps with $O_h$ and $T_d$ symmetry}\news

\subsection{Klein polynomials and the degree $28$ map}

The Skyrme crystal is cubically symmetric so the 
symmetries of rational maps constructed from the crystal are 
related to $O_h$. $O_h$ itself and its subgroup $T_d$, the full
symmetry groups of the cube and tetrahedron, are of particular 
importance.  

To study the interplay of the geometrical method and symmetry, it is 
useful to recall the tetrahedrally symmetric Klein polynomials
\cite{K} and to express our degree $28$ map (\ref{deg28}) in terms of these. In our preferred 
orientation, the tetrahedral vertex and face polynomials $p_+(z)$ and 
$p_-(z)$ are
\begin{align}
p_+(z) &= z^4 + 2 \sqrt{3}i z^2 + 1 \,, \nonumber \\
p_-(z) &= z^4 - 2 \sqrt{3}i z^2 + 1 \,.
\end{align}
The ratio of these degree 4 polynomials is the rational map 
$R(z)=p_+(z)/p_-(z)$ of the $B=4$ Skyrmion. This is not just 
tetrahedrally, but cubically symmetric. The extra symmetry 
under a $90^{\circ}$ rotation sends $z$ to $iz$, and hence $R(z)$ 
to $1/R(z)$. 

\begin{figure}[h!]
\centering
        \begin{subfigure}[b]{0.3\textwidth}
                \centering
		\includegraphics[width=0.95\textwidth]{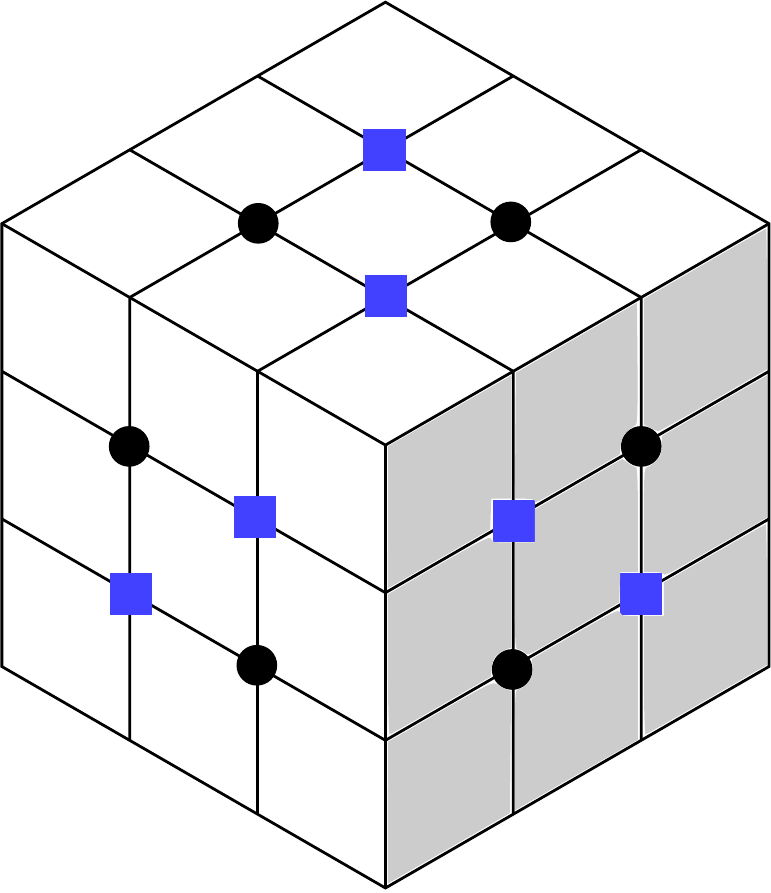}
                \caption{Points on face interiors}
                \label{fig:R_F}
        \end{subfigure}%
        ~ \, 
        \begin{subfigure}[b]{0.3\textwidth}
                \centering
                \includegraphics[width=\textwidth]{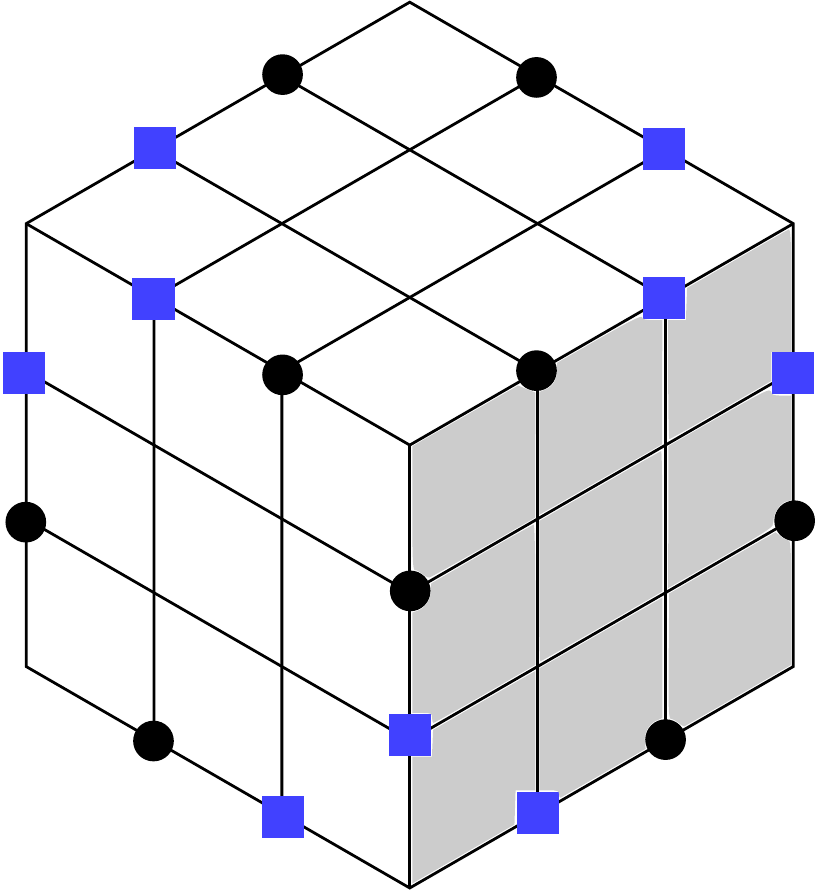}
                \caption{Points on edges}
                \label{fig:R_E}
        \end{subfigure}
        ~ 
        \begin{subfigure}[b]{0.3\textwidth}
                \centering
                \includegraphics[width=\textwidth]{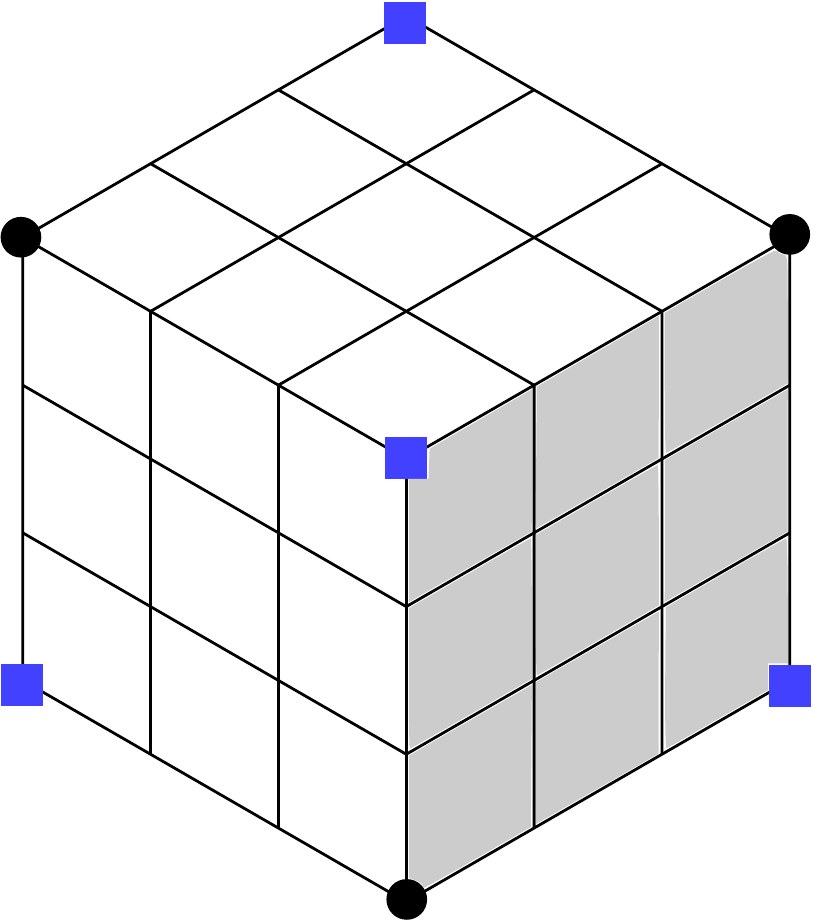}
                \caption{Points on vertices}
                \label{fig:R_v}
        \end{subfigure}
        \caption{Points of the $4 \times 4 \times 4$ cubic grid}
        \label{fig:grid}
\end{figure}

$p_+/p_-$ turns out to be the most important ingredient 
in the geometrically generated rational maps. For example, the
56 points in the outer layer of the $4 \times 4 \times 4$ cubic grid
shown in Figure \ref{fig:cube_grid} can be separated into three subsets:
those on the face interiors, those on the edges and those
on the vertices (see Figure \ref{fig:grid}). Each subset has cubic symmetry.

\begin{figure}[ht]
\centering
\includegraphics[width=0.7\textwidth]{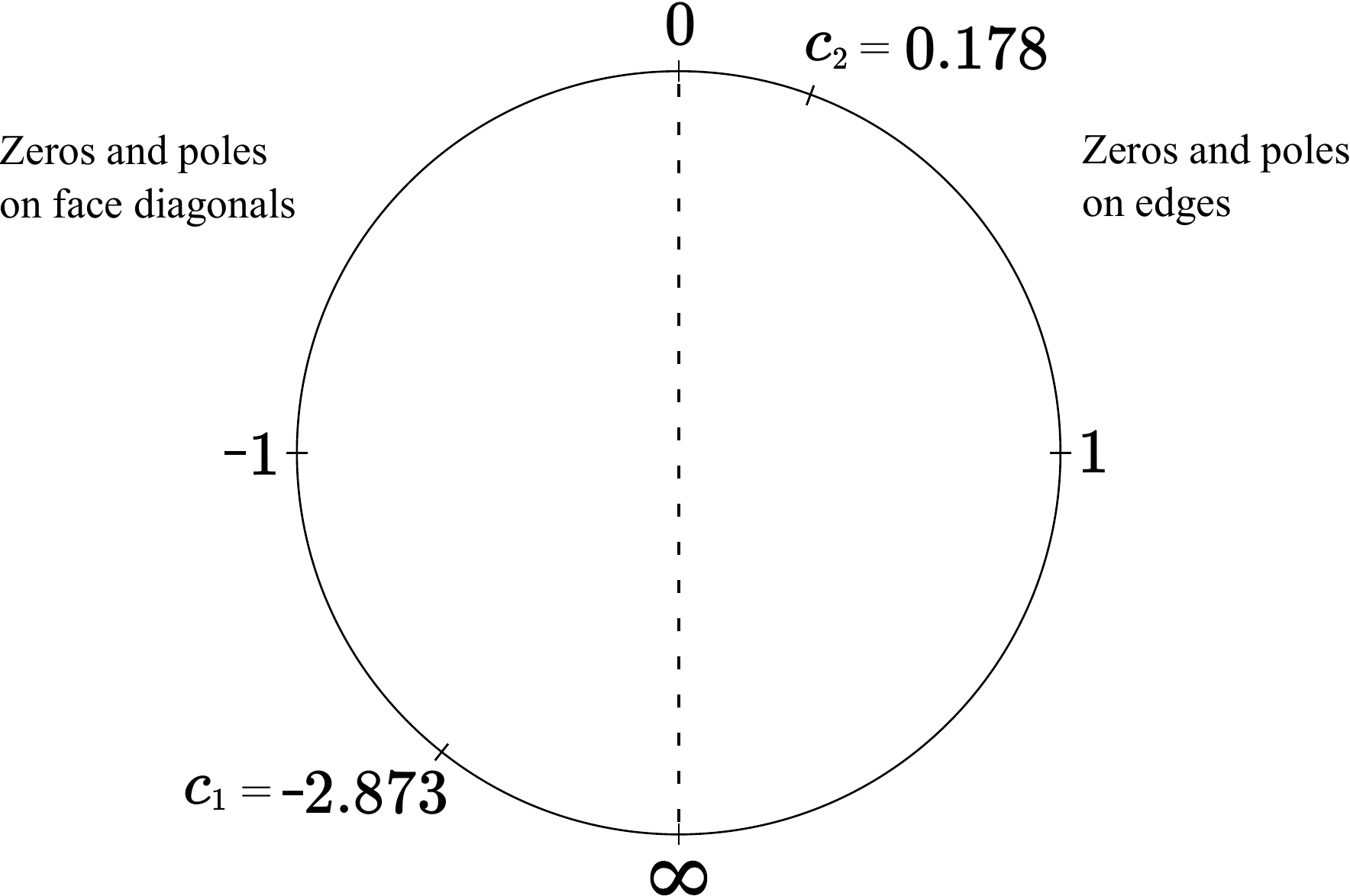}
\caption{Range of $c$}
\label{fig:para_k}
\end{figure}

It is convenient to denote the rational maps constructed from 
these subsets by $R_F$, $R_E$ and $R_V$. They are of 
degrees $12$, $12$ and $4$, respectively, and $R_V$ is the familiar 
cubic degree 4 map (\ref{deg4}). $R_F$ and $R_E$ are (projectively) linearly 
related to ${R_V}^3 = p_+^3/p_-^3$ by 
\be
R_F = \frac{c_1 {p_+}^3 + {p_-}^3}{{p_+}^3 + c_1 {p_-}^3} \,, \quad 
R_E = \frac{c_2 {p_+}^3 + {p_-}^3}{{p_+}^3 + c_2 {p_-}^3} \,,
\label{tetradeg12}
\ee
with $c_1 = -2.873$ and $c_2 = 0.178$. (Both $c_1$ and $c_2$ can be expressed in
analytical form, but this is messy and provides no further
insight.) Relations (\ref{tetradeg12}) can be understood 
by noting that $O_h$ symmetry does not uniquely fix 24 points 
on the Riemann sphere. There is a family with one real parameter $c$ having
this symmetry,
\begin{equation}
R_{(c)} = \frac{c {p_+}^3 + {p_-}^3}{{p_+}^3 + c {p_-}^3} \, .
\label{one_para}
\end{equation}
Starting from $c=-1$, the numerator and the denominator of the rational map $R_{(c)}$ cancel completely and the map takes a constant value of $-1$ which corresponds to the two zeros and the two poles on each face in Figure \ref{fig:R_F} coinciding at the centre of the face. As $c$ varies from $-1$ to $- \infty$, the points move from the face centres along face diagonals to the vertices; the rational map $R_F$ is a special case. At $c=- \infty$, three zeros or three poles coincide at the vertices and ${R_V}^3$ is recovered. We next identify $c=-\infty$ and $c=\infty$ as they give the same rational map. When $c$ varies from $\infty$ to $1$, the points move along the edges and the rational map changes from ${R_V}^3$ to a constant value of $1$ where the zeros and poles coincide at the middle of the edges. The range of the parameter $c$ has now covered the lower half circle of Figure \ref{fig:para_k}. The rational map $R_{(c)}$ behaves similarly for $c$ taking values on the upper half circle but with the numerator and denominator interchanged (zeros become poles and vice versa). The rational map $R_E$ is a special case in this range.

We now see that
\begin{align}
R_V \times R_F \times R_E &= \frac{p_+}{p_-} \left( \frac{c_1 p_+^3 + p_-^3}{p_+^3 + c_1 p_-^3} \right) \left( \frac{c_2 p_+^3 + p_-^3}{p_+^3 + c_2 p_-^3} \right) \nonumber \\
&= \frac{p_+}{p_-} \left( \frac{c_1 c_2 \ p_+^6 + (c_1 + c_2) p_+^3 p_-^3 +
    p_-^6}{p_+^6 + (c_1 + c_2) p_+^3 p_-^3 + c_1 c_2 \ p_-^6} \right) \,, 
\label{deg28_geo}
\end{align}
and this can be compared with the degree $28$ map (modulo sign flips) constructed by 
the optimisation method,
\begin{equation}
R = \frac{p_+}{p_-} \left(\frac{-C p_+^6 - D p_+^3 p_-^3 + p_-^6}
{p_+^6 - D p_+^3 p_-^3 - C p_-^6}\right) \,. 
\label{deg28_old}
\end{equation}

The values of $C$ and $D$ calculated from $c_1$ and $c_2$ are $0.51$ and $2.70$, respectively, 
which are moderately close to the values $C=0.33$ and $D=1.64$ in \cite{BMS}.

\subsection{$B=20$ Skyrmions}

We have also used the rational maps $R_F$, $R_E$ and $R_V$ to seek 
the $B=20$ Skyrmion, which for massive pions was not previously 
firmly established. The $\alpha$-particle model of nuclei suggests that the 
$B=20$ Skyrmion is formed from five $B=4$ Skyrmions. A solution of this type was found earlier \cite{BMS}, but
was not of the expected triangular bi-pyramid shape, had little
symmetry and was probably not of minimal energy. 

\begin{figure}[ht]
	\centering
        \begin{subfigure}[b]{0.35\textwidth}
                \centering
                \includegraphics[width=\textwidth]{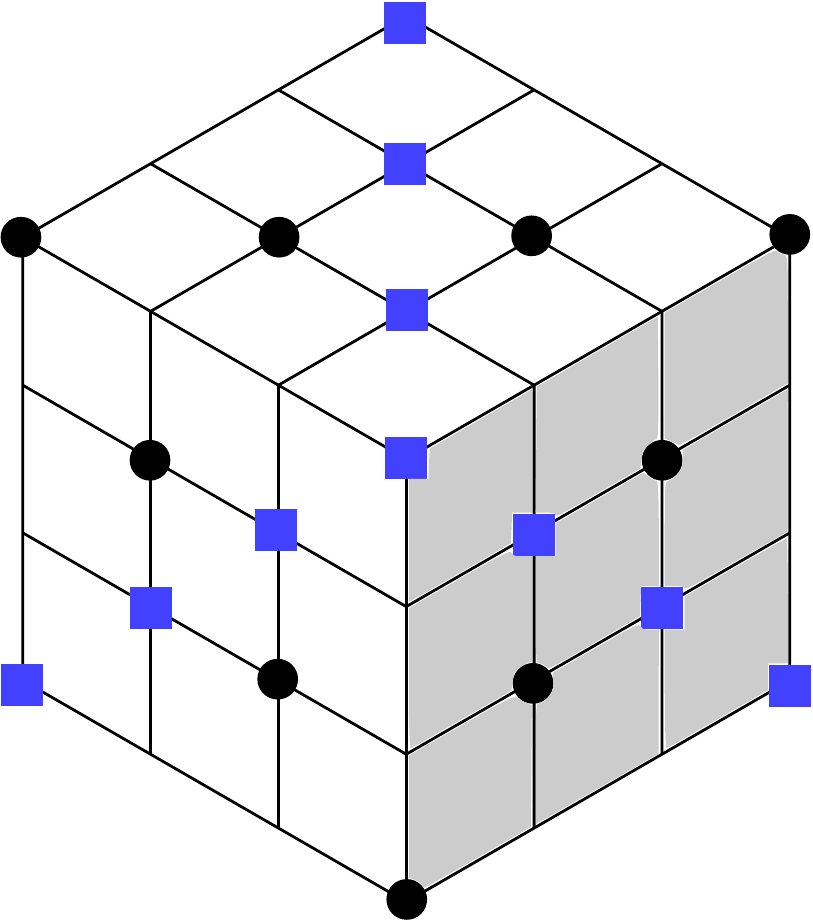}
                \caption{Points generating the $T_d$-symmetric Skyrmion}
                \label{fig:B_20_TCube}
        \end{subfigure}
        \qquad 
        \begin{subfigure}[b]{0.35\textwidth}
                \centering
                \includegraphics[width=\textwidth]{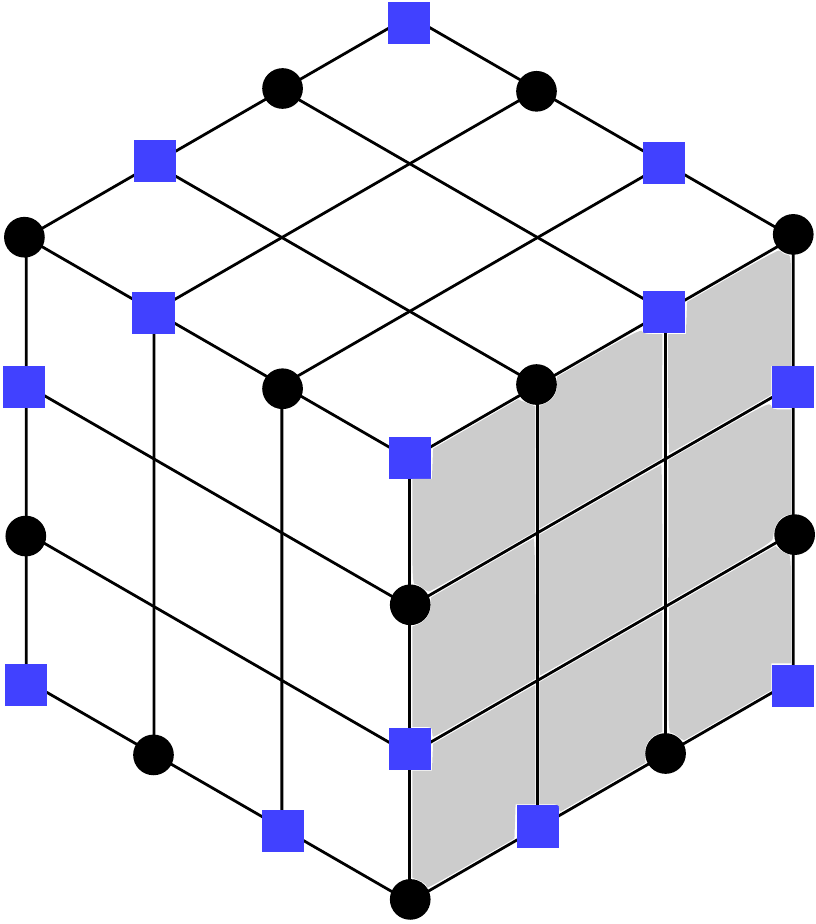}
                \caption{Points generating the $D_{2h}$-symmetric Skyrmion}
                \label{fig:B_20_D2hCube}
        \end{subfigure}
        \caption{Grid points used for $B=20$}
        \label{fig:B_20_Cube}
\end{figure}

Two useful rational maps are $R= R_V \times R_F$ and
$R = R_V \times R_E$. Their zeros and poles are shown in 
Figure \ref{fig:B_20_Cube}.
Each has degree $16$ and can be used as an outer map. The inner map is
the cubically symmetric degree $4$ map (\ref{deg4}). Using these
pairs of maps in the double rational map ansatz, and relaxing, gives
two candidate $B=20$ Skyrmions, but neither has a bi-pyramid shape. 
The cubic symmetry is also not rigorously enforced by the numerics, and it 
ends up broken. The first Skyrmion has $T_d$ symmetry, and the
second only $D_{2h}$ symmetry.

\begin{figure}[h!]
	\centering
        \begin{subfigure}[b]{0.35\textwidth}
                \centering
                \includegraphics[width=\textwidth]{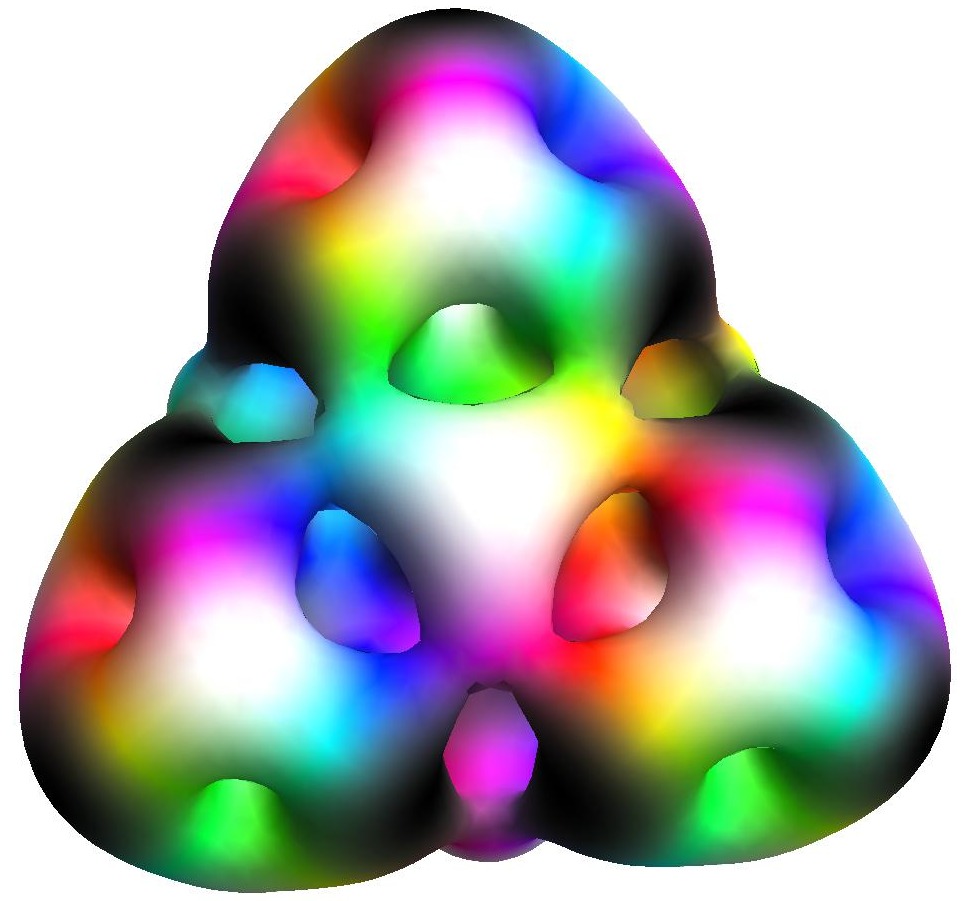}
                \caption{$B=20$ Skyrmion with $T_d$ symmetry \newline}
                \label{fig:B_20_TdCube}
        \end{subfigure}
        \qquad 
        \begin{subfigure}[b]{0.35\textwidth}
                \centering
                \includegraphics[width=\textwidth]{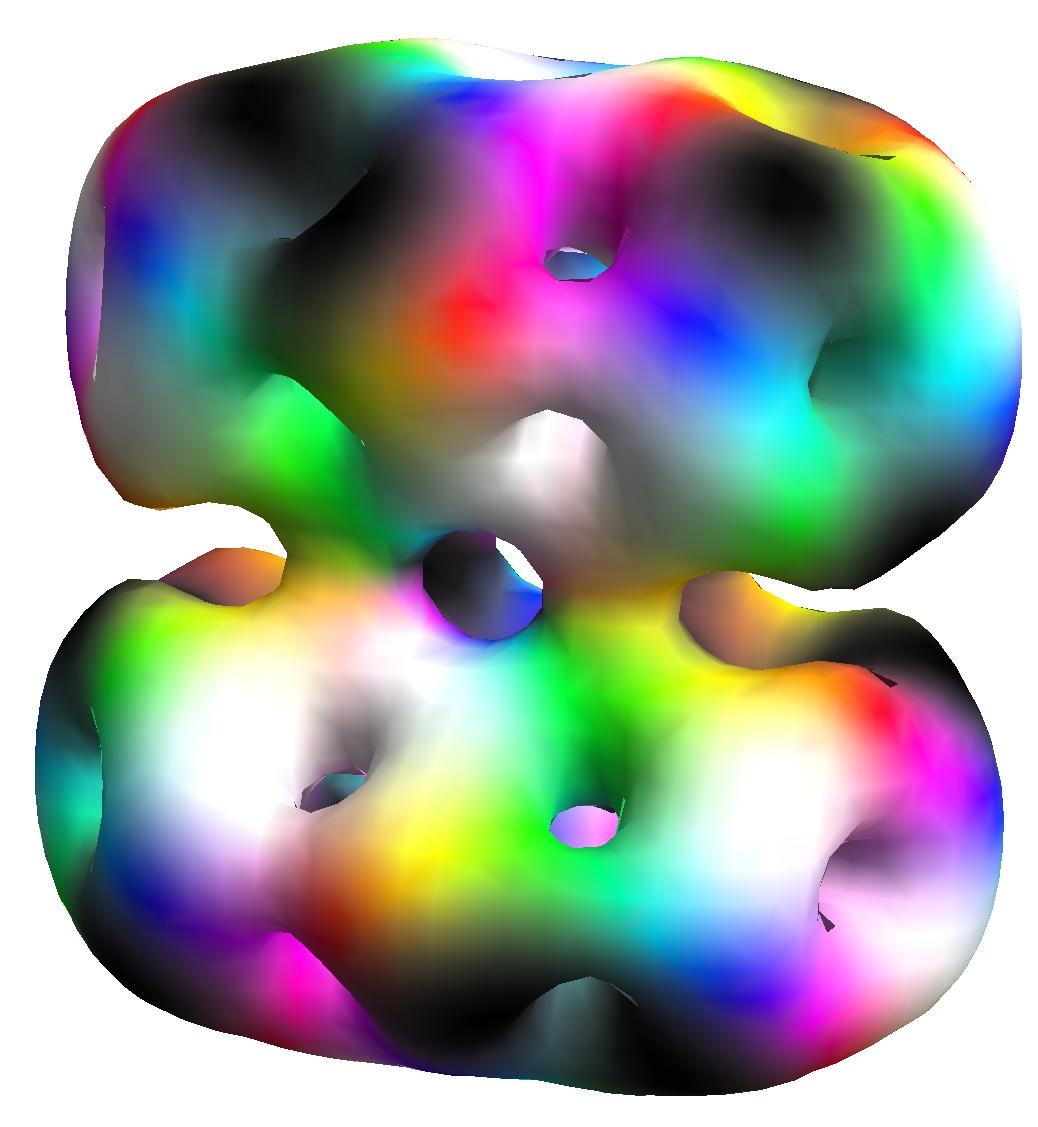}
                \caption{$B=20$ Skyrmion formed by stacking two $B=10$ Skyrmions}
                \label{fig:B_20_D2h}
        \end{subfigure}
        \caption{$B=20$ Skyrmions}
\end{figure}

The $T_d$-symmetric solution has 
an energy per baryon $E/B = 1.277$. It can be interpreted as four slightly 
distorted $B=4$ cubes at the vertices of a tetrahedron and four 
$B=1$ Skyrmions at the face centres of the tetrahedron (see
Figure \ref{fig:B_20_TdCube}). Each $B=1$ Skyrmion is oriented such
that the colour equator ($\hat {\bpi}_1$--$\hat {\bpi}_2$ plane) is 
parallel to the face of the tetrahedron containing the Skyrmion. 
Its primary colours match the colours of the faces of the 
$B=4$ cubes that it touches. As a result, the $B=1$ and $B=4$ 
Skyrmions are all attracting. The $D_{2h}$-symmetric solution 
has slightly lower energy per baryon, $E/B = 1.274$, and appears to be the
lowest energy solution for $B=20$. The Skyrmion consists of two
loosely touching clusters, each in the form of the known $B=10$ 
Skyrmion (see Figure \ref{fig:B_20_D2h})\cite{BS11}. 

\begin{figure}[ht]
\centering
\includegraphics[width=5cm]{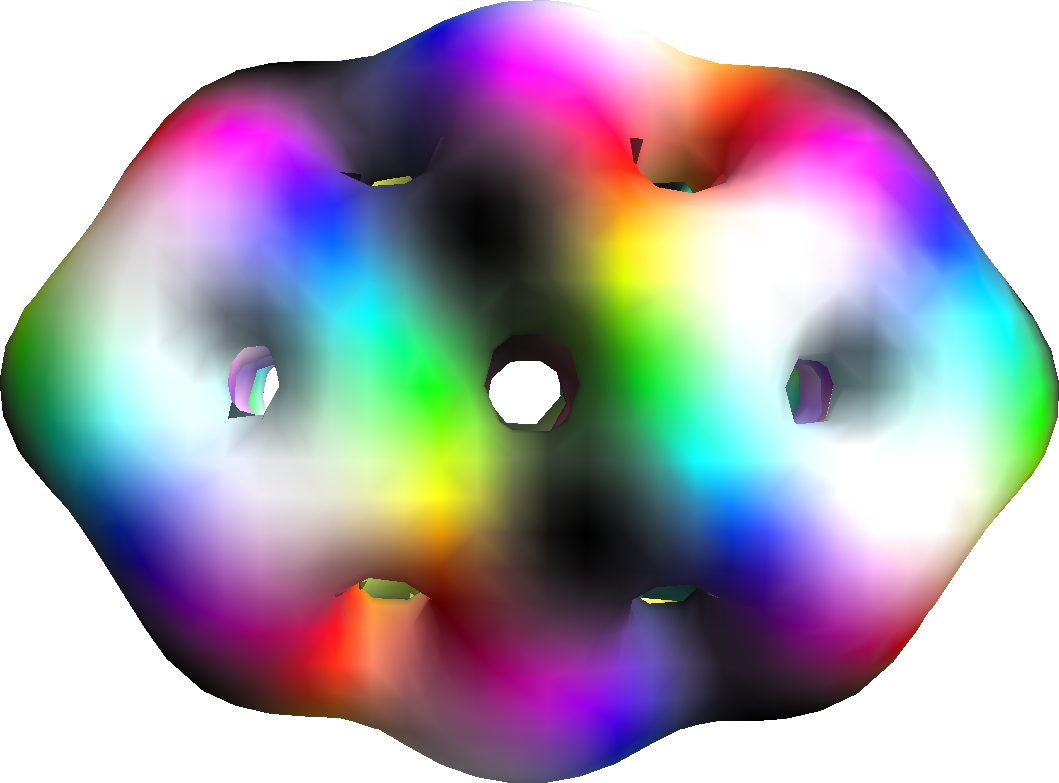}
\caption{Top view of $B=20$ with $D_{2h}$ symmetry}
\label{fig:B_20_D2h_Top}
\end{figure}

The $B=10$ Skyrmion itself resembles two $B=4$ cubes bound together by 
two $B=1$ Skyrmions. Figure \ref{fig:B_20_D2h_Top} shows the top view 
of the $B=20$ Skyrmion which looks the same as the $B=10$ Skyrmion 
in \cite{BS11}. The $B=10$ Skyrmion can be reproduced using the geometric method, and has 
$E/B = 1.280$. As expected, the $B=20$ Skyrmion has slightly 
lower energy than two well-separated $B=10$ Skyrmions.

Since the energy difference between the $T_d$-symmetric and $D_{2h}$-symmetric 
solutions is only about $0.2\%$, it is reasonable to ask how similar
the solutions are. The relaxation of the Skyrme field was monitored. It
appears that for the $T_d$-symmetric Skyrmion, it is energetically favourable 
to rotate one of the edges of the tetrahedron relative to the opposite
one by $90^\circ$. If a deformation is introduced
breaking the tetrahedral symmetry, the field relaxes to 
the $D_{2h}$-symmetric solution.

\begin{figure}[ht]
\centering
\includegraphics[width=4.5cm]{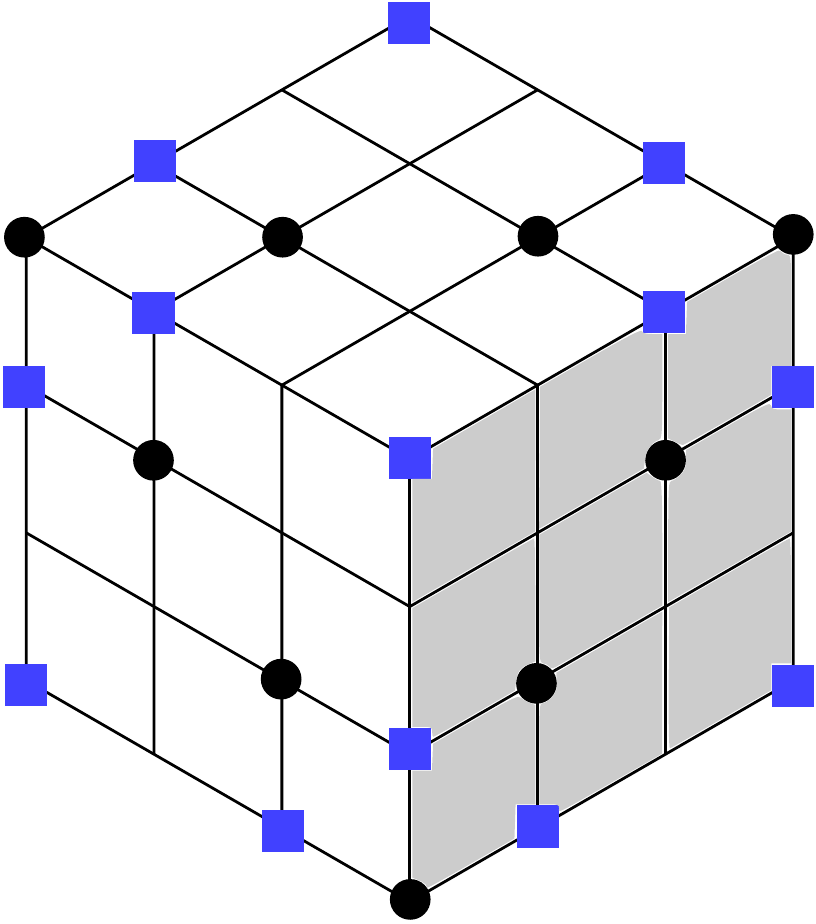}
\caption{Zeros and poles with $T_d$ Symmetry}
\label{fig:Td_Grid}
\end{figure}

To obtain the $T_d$-symmetric solution, it is actually preferable
to start with an outer rational map with $T_d$ rather than
$O_h$ symmetry. The initial field configuration will likely be
closer to the Skyrmion and convergence faster. A suitable degree $16$ outer map has 
been constructed using the set of points shown in 
Figure \ref{fig:Td_Grid}. The degree $4$ inner map is unchanged. This
combination builds in the four $B=4$ Skyrmions and the four $B=1$
Skyrmions rather effectively.

This rational map with $T_d$ symmetry can again be written in terms 
of $p_+$ and $p_-$, and is
\begin{equation}
R = \left( \frac{1+c_2}{1+c_1} \right) \frac{p_+}{p_-} 
\left(\frac{c_1 p_+^3 + p_-^3}{p_+^3 + c_2 p_-^3}\right) \, ,
\end{equation}
where $c_1$ and $c_2$ are as in (\ref{tetradeg12}). It is a bit more complicated than an $O_h$-symmetric map.

\section{Skyrmions obtained by other methods}\news

It is established that Skyrmions often cluster into $B=4$ units, 
even when this is not imposed in the first place. 
One may therefore take the $B=4$ Skyrmion as a basic building block
and assemble larger Skyrmions from it, using the product ansatz
followed by numerical relaxation. This has been done previously for 
$B=8$, $B=12$ and $B=32$. We have now found solutions for $B=24$ in this way.

\begin{figure}[ht]
\centering
\includegraphics[width=0.26\textwidth]{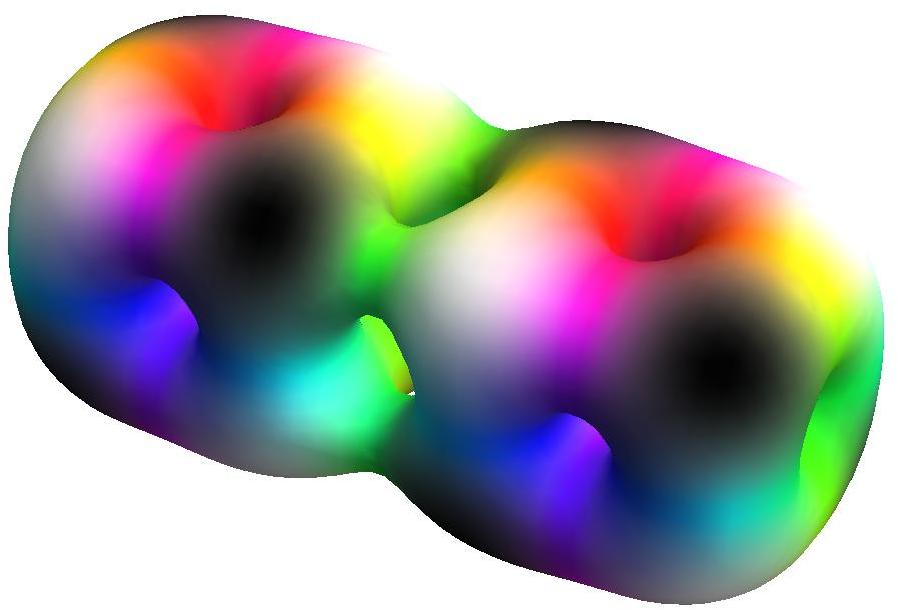}\hspace{8mm}
\includegraphics[width=0.26\textwidth]{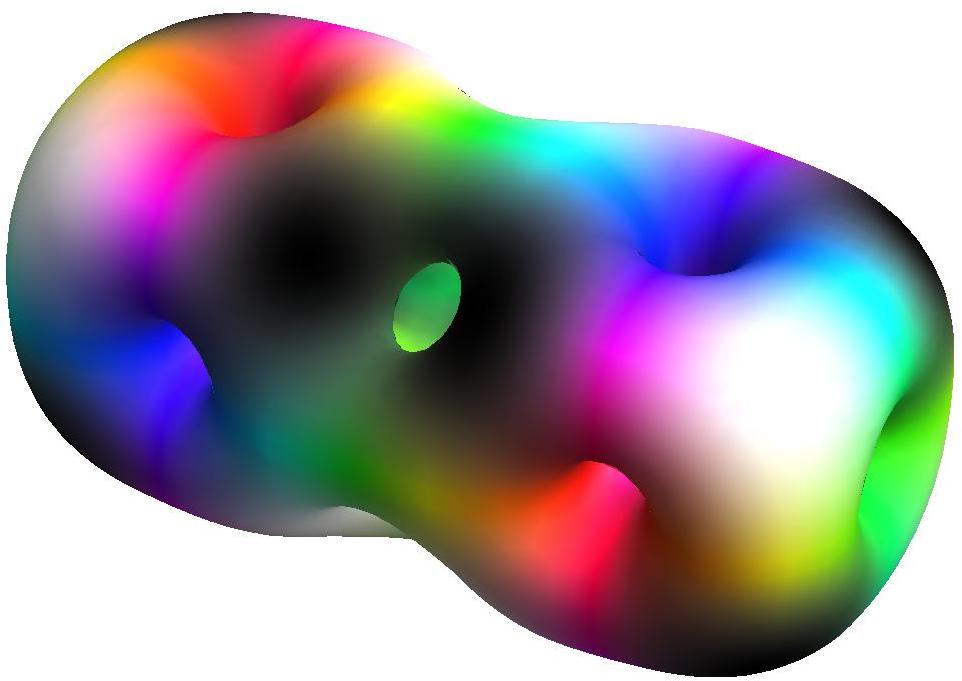}\hspace{8mm}
\includegraphics[width=0.3\textwidth]{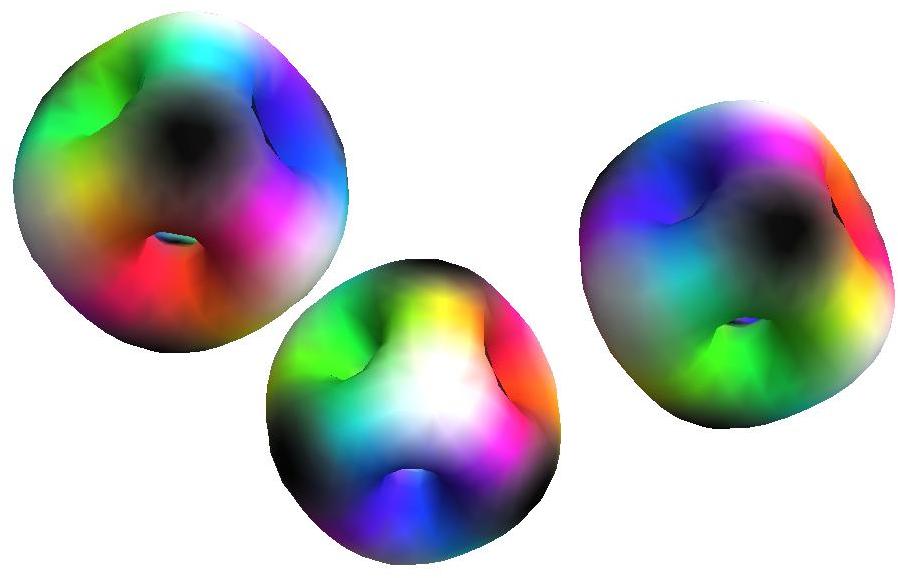}
\caption{Left: Two $B=4$ cubes bound together into a $B=8$ solution. Middle: The same, but one cube is rotated around the axis of separation; this is the lower energy solution. Right: Three cubes in an L-shape.}
\label{fig:B_8_12}
\end{figure}

The cubic $B=4$ Skyrmion (Figure \ref{fig:B_4}) has
alternating white and black half-Skyrmions at the corners. In the
standard orientation defined by the rational map (\ref{deg4}), its face 
colours are red, green and blue, and opposite faces have the same colour.
This means that two cubes in the same orientation, face-to-face, will
have matching face colours, but the corner colours will not 
match. This is the situation in the Skyrme crystal, where the 
$B=4$ cubes are all oriented the same way, and also in crystal chunks, such 
as the $B=32$ and $B=108$ Skyrmions.

Given just two cubes, we can twist one of them by $90^\circ$ around
the axis of separation (see Figure \ref{fig:B_8_12}). Then both the face and corner colours
match, although the edge colours do not. This is the minimal energy 
configuration in the $B=8$ sector for the Skyrme model with massive 
pions. However, it is not possible to construct a crystal with these twists: 
After arranging three $B=4$ cubes in an L-shape with $90^\circ$ 
twists, the neighbouring empty space will be bounded by two faces 
of the same colour, so a further cube cannot be inserted with 
low energy to make a $B=16$ square (see Figure \ref{fig:B_8_12}, right).

\begin{figure}[h!]
	\centering
        \begin{subfigure}[b]{0.35\textwidth}
                \centering
                \includegraphics[width=\textwidth]{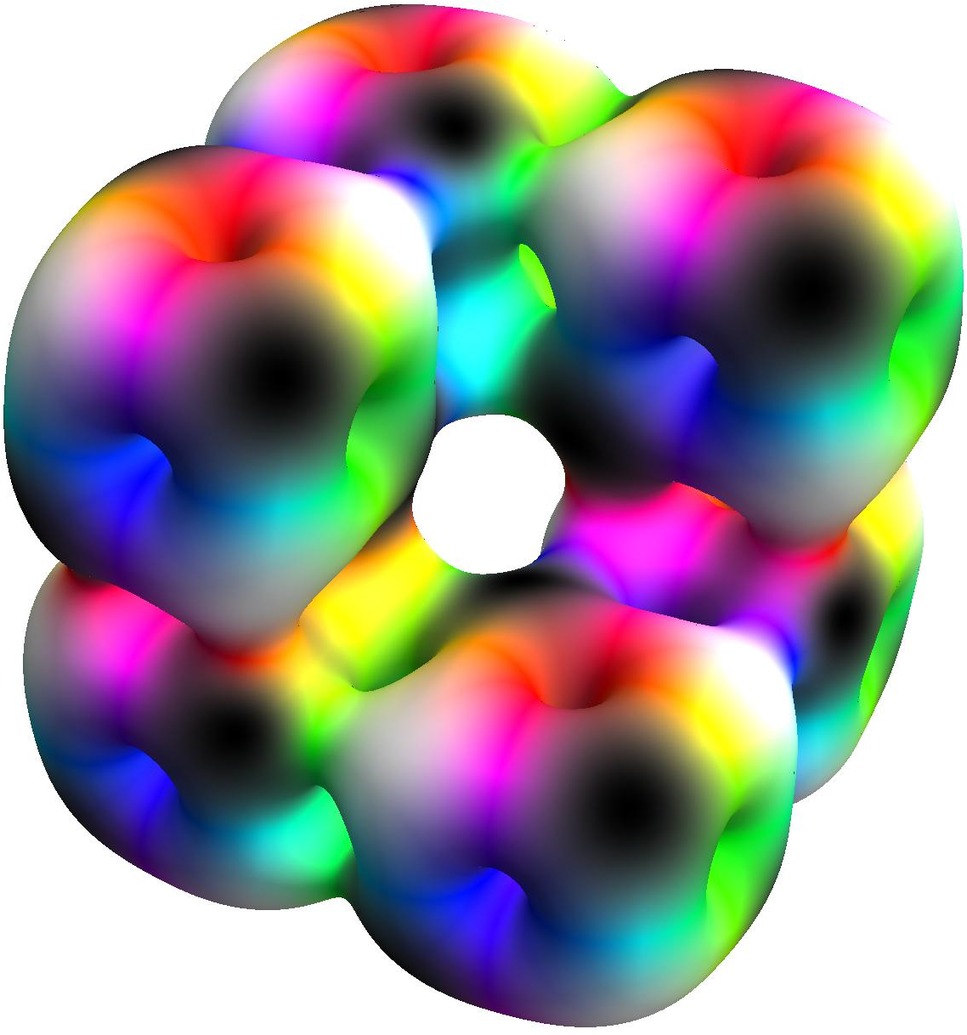}
                \caption{$B=24$ Skyrmion as crystal chunk}
                \label{fig:B_24_1}
        \end{subfigure}
        \qquad 
        \begin{subfigure}[b]{0.35\textwidth}
                \centering
                \includegraphics[width=\textwidth]{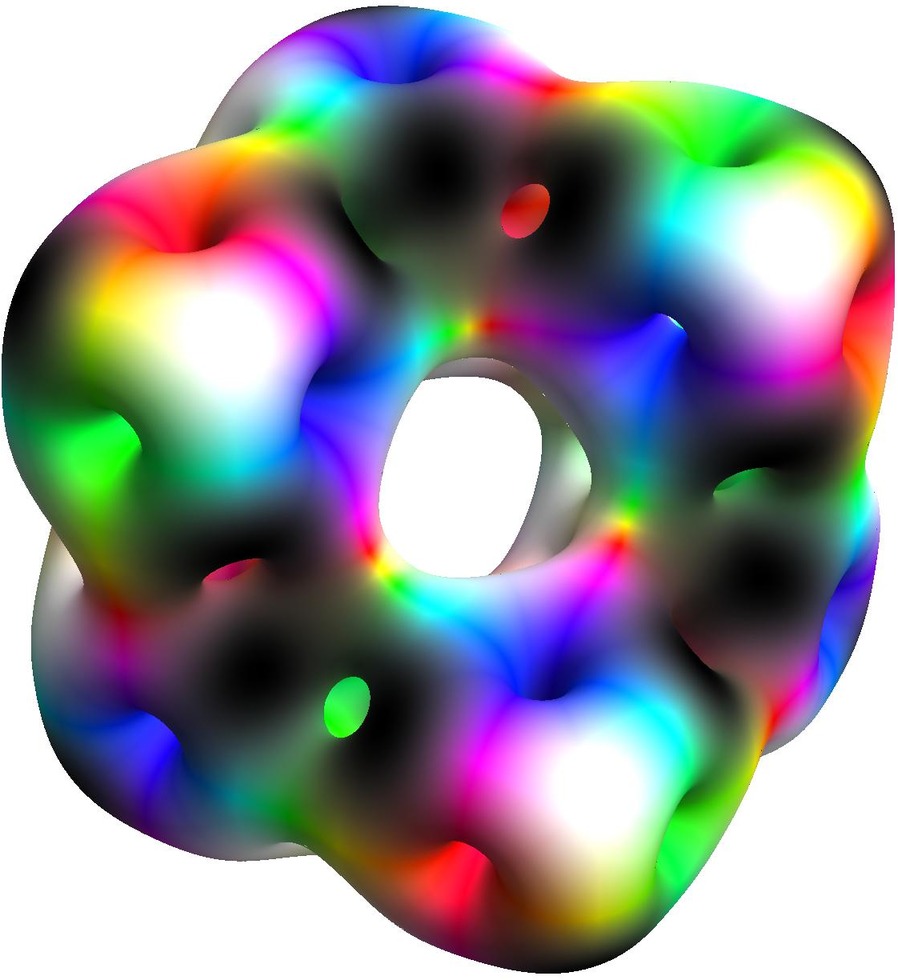}
                \caption{$B=24$ Skyrmion with twisted cubes}
                \label{fig:B_24_2}
        \end{subfigure}
        \caption{$B=24$ Skyrmions}
\end{figure}

One can get a $B=24$ solution, as a crystal chunk, by removing two 
$B=4$ cubes from opposite corners of the $B=32$ Skyrmion. This results 
in a non-planar ring of six $B=4$ cubes, all
with the same orientation (Figure \ref{fig:B_24_1}). However, 
in this case, we can do better. Because of the two missing corner 
cubes, we can re-orient the remaining six cubes so that 
each neighbouring pair has a $90^\circ$ relative twist around 
its separation axis (Figure \ref{fig:B_24_2}). This results 
in a lower energy. These two $B=24$ solutions have $E/B = 1.273$ and
$E/B = 1.269$, respectively, and the latter may be the true
$B=24$ Skyrmion of minimal energy.

\section{Conclusions}\news

In this paper we have presented several new Skyrmions for pion mass 
parameter $m=1$, with baryon numbers as high as $B=108$. We have 
used insight obtained from the infinite Skyrme crystal to develop 
a geometrical method to construct multi-layer rational map ans\"atze 
for the Skyrme field. This gives 
fairly simple algebraic expressions for initial data, which have 
then been relaxed numerically to obtain Skyrmions which are either 
global or local minima of the energy. 

The cubic $B=32$ Skyrmion, previously known, was constructed afresh using a
double rational map ansatz. A range of novel Skyrmions from $B=24$ up 
to $B=31$, related to this $B=32$ Skyrmion, were then obtained by 
modifying the outer rational map, so as to remove one unit of baryon 
number from any of the eight corners. Further modifications 
of the rational map led to new Skyrmions with $B=20$. 

For the first time, a three-layer rational map ansatz has been 
used. This successfully generated the cubic $B=108$ Skyrmion, 
the first Skyrmion with baryon number $B > 100$ to be found. Corner 
cutting gave solutions in the range $B=100$ to $B=107$. 

Using another approach, we have found what may be the lowest energy
Skyrmion with $B=24$. This was constructed by gluing together, with 
twists, six $B=4$ cubic Skyrmions into a ring, and relaxing.

The next step would be to consider quantisation of the newly
found Skyrmions in order to get further nuclear spectra from
the Skyrme model. Nuclear spectra calculations are currently limited
to baryon numbers no higher than $16$ \cite{BMSW,Wood}.

\begin{table}
\begin{center}
\begin{tabular}{ | c | c | c | c | p{6cm} |}
\hline
Baryon Number $B$ & Energy $E$ & $E/B$ & Symmetry & Comment\\ \hline
$1$ & $1.465$ & $1.465$ & $O(3)$ & Hedgehog \\ \hline
$2$ & $2.77$ & $1.385$ & $D_{\infty h}$ & Toroid \\ \hline
$3$ & $4.02$ & $1.340$ & $T_d$ & Tetrahedral Skyrmion \\ \hline
$4$ & $5.18$ & $1.295$ & $O_h$ & Cubic Skyrmion \\ \hline
$8$ & $10.25$ & $1.281$ & $D_{4h}$ & Two $B=4$ cubes with $90^\circ$ twist \\
$ $ & $10.28$ & $1.285$ & $D_{4h}$ & Two $B=4$ cubes without twist \\ \hline
$10$ & $12.80$ & $1.280$ & $D_{2h}$ & \\ \hline
$20$ & $25.47$ & $1.274$ & $D_{2h}$ & Two $B=10$ clusters  \\
$ $ & $25.53$ & $1.277$ & $T_d$ & \\ \hline
$24$ & $30.47$ & $1.269$ & $D_{3d}$ & Six $B=4$ cubes with twists \\
$ $ & $30.57$ & $1.273$ & $D_{3d}$ & Six $B=4$ cubes without twists \\ 
$ $ & $30.80$ & $1.283$ & $O_h$ & Octahedral cluster of $B=4$ cubes dual to $B=32$ cubic Skyrmion \\ \hline
$25$ & $31.79$ & $1.272$ & $C_{1h}$ & A cluster of $B=4$ and $B=3$ Skyrmions \\ \hline
$26$ & $33.05$ & $1.271$ & $C_{2h}$ & Two $B=13$ clusters \\ \hline
$27$ & $34.39$ & $1.274$ & $C_{3v}$ & \\ \hline
$28$ & $35.57$ & $1.270$ & $T_d$ & \\ \hline
$29$ & $36.78$ & $1.268$ & $C_{3v}$ & \\ \hline
$30$ & $38.00$ & $1.267$ & $D_{3d}$ & \\ \hline
$31$ & $39.25$ & $1.266$ & $C_{3v}$ & \\ \hline
$32$ & $40.51$ & $1.266$ & $O_h$ & Cubic cluster of eight $B=4$ cubes \\ \hline
$100$ & $125.68$ & $1.257$ & $O_h$ & \\ \hline
$101$ & $126.86$ & $1.256$ & $C_{3v}$ & \\ \hline
$102$ & $128.05$ & $1.255$ & $D_{3d}$ & \\ \hline 
$103$ & $129.26$ & $1.255$ & $C_{3v}$ & \\ \hline
$104$ & $130.47$ & $1.255$ & $T_d$ & \\ \hline
$105$ & $131.71$ & $1.254$ & $C_{3v}$ & \\ \hline
$106$ & $132.95$ & $1.254$ & $D_{3d}$ & \\ \hline
$107$ & $134.21$ & $1.254$ & $C_{3v}$ & \\ \hline 
$108$ & $135.47$ & $1.254$ & $O_h$ & Cubic cluster of 27 $B=4$ cubes \\ \hline
$\infty$ & --- & $1.238$ & $O_h$ & Skyrme Crystal \cite{Fei} \\
\hline 
\end{tabular}
\end{center}
\caption{Energies and symmetries of Skyrmions (for $m=1$). The energy $E$ is accurate to $\pm0.01$}
\label{tab:energy}
\end{table}

\newpage

\section*{Acknowledgements}

D.T.J Feist is supported by the Gates Cambridge Trust and EPSRC. P.H.C. Lau is supported by Trinity College, Cambridge.

\appendix
\section{Appendix: Numerical methods}\news

All of the Skyrme field relaxations were done with the numerical 
methods developed in \cite{Fei}, with some extensions. The 
numerics are based on an $(N+1)^3$ cubic grid with lattice constant 
$h$, on which the Skyrme energy from equation (\ref{skyenergy}) 
was discretised using sixth-order finite differences. This particular 
discretisation order was chosen because it gives the highest precision 
per computation time.

To find the minimum of the discretised energy, a nonlinear conjugate
gradient (NLCG) method was chosen \cite{NLCG}. NLCG can be seen as a
geometrically enhanced version of gradient descent. The latter is a
very slow method, in particular when the Hessian is of poor
condition. NLCG converges much faster.

Another way to find minima of the energy is to compute the
time development using a version of the dynamical Skyrme field 
equations. The advantage is that the field will accelerate towards 
an energy minimum, but the disadvantage is that it will then overshoot if no 
measures are taken. The simplest remedy is to take out kinetic 
energy by introducing a friction term, but this eliminates the 
real advantage over gradient descent. A more advanced method is 
to take out all kinetic energy whenever the potential energy is 
increasing. 

One may ignore nonlinear terms with time derivatives in the
Skyrme field equation in this process, which gives a simpler, 
if not quite correct time development. This does not matter, as long as 
the Lagrangian includes the correct static energy. The time development will still converge to minima of the
energy \cite{BBT}. This method gives rapid convergence and has
been used to compute many Skyrmions \cite{BS3a}.

Starting close to an energy minimum, the NLCG method converges
sufficiently rapidly, in our experience. From further away, time 
development is often faster at producing the required structural 
changes to the field. This is expected, because far away from a 
quadratic minimum, NLCG is not much better than simple gradient 
descent methods, and these tend to zig-zag.

All numerical methods were implemented in C with thread-level
parallelisation for computing gradients. Initial field configurations 
were generated using python scripts. Usually, an initial field 
configuration was computed on an $N=40$ grid with lattice 
spacing $h=0.2$. This makes rapid exploration of the solution 
space possible; the computation time to generate relaxed solutions 
on current Desktop PCs is a few hours.
One should make sure all major relaxation is finished and no 
structural changes are going on before making further 
refinements to improve energy estimates. 

A common method to allow for lattice effects is to correct energies 
by a factor of $B/B_{\rm num}$, so that
\be
E = \frac{B}{B_{\rm num}} E_{\rm num} \,,
\ee
where $E_{\rm num}$ is the numerical energy, and $B_{\rm num}$ is the 
baryon number obtained as the numerically evaluated integral of the 
baryon density (\ref{eq:bardens}) using the same discretisation scheme as
for the energy. $B$ is the (true) integer baryon number and $E$ is the
corrected energy estimate. This method was often used to estimate
Skyrmion energies \cite{BS3a}, and works surprisingly well in the
massless pion case, because the numerical calculations appear to
underestimate both the energy and the baryon number by a very similar
factor. However, it does not work well for Skyrmions with massive 
pions. The reason is that the pion mass contribution to the energy 
is not similarly underestimated by numerics, probably because it 
does not contain any derivatives. One could try to develop a new 
extrapolation method, separating derivative and non-derivative 
energy contributions, but we haven't tried this.

Instead, to get accurate Skyrmion energies, we gradually increased $N$ 
and gradually decreased $h$, so that the effects of both could be
estimated, and an extrapolation to the continuum limit made. This
gives both a good energy estimate and an idea of the precision. The Skyrmion energies are presented in Table \ref{tab:energy}; the accuracy is $\pm 0.01$.
This method works well for massive pions, because the Skyrme field
decays quickly (exponentially as $\exp(-m|\mathbf x|)$, with $m=1$ 
here), but would be less useful for massless pion computations, as 
the Skyrme field decays only algebraically.

\end{document}